\documentclass[final,3p,12pt]{elsarticle}
\usepackage{amssymb}
\usepackage[dvipsnames]{xcolor}
\usepackage{multirow}
\usepackage{amsmath}
\usepackage{amsfonts}
\usepackage{amssymb}
\usepackage{hyperref}

\makeatletter
\def\ps@pprintTitle{%
	\let\@oddhead\@empty
	\let\@evenhead\@empty
	\def\@oddfoot{}%
	\let\@evenfoot\@oddfoot}
\makeatother

\begin{document}

\begin{frontmatter}

\author[add1]{Paraskevi Nousi\corref{cor1}}
\ead{paranous@csd.auth.gr}

\author[add1]{Avraam Tsantekidis\corref{cor1}}
\ead{avraamt@csd.auth.gr}

\author[add1]{Nikolaos Passalis}
\ead{passalis@csd.auth.gr}

\author[add2]{Adamantios Ntakaris}
\ead{adamantios.ntakaris@tut.fi}

\author[add3]{Juho Kanniainen}
\ead{juho.kanniainen@tut.fi}

\author[add1]{Anastasios Tefas}
\ead{tefas@aiia.csd.auth.gr}

\author[add2]{Moncef Gabbouj}
\ead{moncef.gabbouj@tut.fi}

\author[add2,add4]{Alexandros Iosifidis}
\ead{alexandros.iosifidis@eng.au.dk}

\address[add1]{Department of Informatics, Aristotle University of Thessaloniki, Greece}
\address[add2]{Laboratory of Signal Processing, Tampere University of Technology, Tampere, Finland}
\address[add3]{Laboratory of Industrial and Information Management, Tampere University of Technology, Tampere, Finland}
\address[add4]{Department of Engineering, Electrical and Computer Engineering, Aarhus University, Denmark}

\cortext[cor1]{Corresponding authors; Equal contribution}

\title{Machine Learning for Forecasting Mid Price Movement using Limit Order Book Data}

\begin{abstract}
Forecasting the movements of stock prices is one the most challenging problems in financial markets analysis. In this paper, we use Machine Learning (ML) algorithms for the prediction of future price movements using limit order book data. Two different sets of features are combined and evaluated: handcrafted features based on the raw order book data and features extracted by ML algorithms, resulting in feature vectors with highly variant dimensionalities. Three classifiers are evaluated using combinations of these sets of features on two different evaluation setups and three prediction scenarios. Even though the large scale and high frequency nature of the limit order book poses several challenges, the scope of the conducted experiments and the significance of the experimental results indicate that Machine Learning highly befits this task carving the path towards future research in this field.
\end{abstract}


\end{frontmatter}

\section{Introduction}
\label{sec:introduction}
Forecasting of financial time series is a very challenging problem and has attracted scientific interest in the past few decades. Due to the inherently noisy and non-stationary nature of financial time series, statistical models are unsuitable for the task of modeling and forecasting such data. Thus, the nature of financial data necessitates the utilization of more sophisticated methods, capable of modeling complex non-linear relationships between data, such as Machine Learning (ML) algorithms.

Early Machine Learning approaches to this problem included shallow Neural Networks (NNs) \cite{kaastra1995forecasting,kaastra1996designing}, and Support Vector Machines (SVMs) \cite{tay2001application,cao2003support,lu2009financial}. However, the lack of appropriate training and regularization algorithms for Neural Networks at the time, such as the dropout technique \cite{srivastava2014dropout}, rendered them susceptible to over fitting the training data. Support Vector Machines were deemed as better candidates for this task, as their solution implicitly involves the generalization error.

The development of effective and efficient training algorithms for deeper architectures \cite{glorot2010understanding}, in conjunction with the improved results such models presented, steered scientific interests towards Deep Learning techniques in many domains. Deep Learning methods are capable of modeling highly non-linear, very complex data, making them suitable for application to financial data \cite{langkvist2014review}, as well as time series forecasting \cite{rahman2016layered}.

Furthermore, ML techniques which perform feature extraction may uncover robust features, better-suited to the specific task at hand. Autoencoders \cite{vincent2008extracting}, are Neural Networks which learn new features extracted from the original input space, which can be used to enhance the performance of various tasks, such as classification or regression. Bag-of-Features (BoF) models comprise another feature extraction method that can be used to extract representations of objects described by multiple feature vectors, such as time-series \cite{baydogan2013bag,iosifidis2013multidimensional}. The process of feature extraction is of major importance as it can significantly affect the performance of the used machine learning algorithms.

In this paper we utilize various Machine Learning algorithms and data preprocessing techniques for the prediction of future price movements of stocks. 
The dataset we use comes from the temporal progression of the limit order book (LOB) which is the highest resolution possible one can observe the stock markets in. The characteristics of LOB data are discussed in detail in Section \ref{sec:handcrafted}, along with a comprehensive description of expertly hand-crafted features that can be extracted from a LOB.

The main contribution of this paper is a very extensive study into the significance of the information provided by the limit order book for the task of predicting future mid price movements of stocks. Handcrafted features, founded on financial expertise, and features extracted by sophisticated Machine Learning techniques are combined and utilized towards this purpose. Three scenarios are assessed regarding the span of time for which predictions are made: a scenario where the movement of the mid price of the immediately succeeding sample in time is predicted, one where the average mid price movement of the next five samples is predicted and, last, one where the average movement of the mid price for the next ten samples is predicted. Multiple ML classifiers are evaluated on two experimental setups for all three scenarios and for various combinations of the extracted features, taking into consideration the class imbalance which accompanies the data as well as the real-time requirements of High Frequency Trading. The results achieved indicate that the LOB contains valuable information which, in conjunction with various Machine Learning algorithms, can give meaningful insight into the stock market trend and lead the models to make accurate predictions, without any external human intervention. Although Machine Learning techniques have been widely used to model other types of financial data \cite{park2015using,yu2008forecasting}, only recently have they begun to be applied and evaluated on LOB data \cite{kercheval2015modelling}.

The rest of the paper is organized as follows. Section~\ref{sec:related_work} presents previous related work upon which we build. A description of the data contained in a limit order book and used for this work is presented in Section~\ref{sec:LOB}. The ML algorithms used to extract features from the original financial data are described in Section~\ref{sec:proposed_methodology}, followed by the classification algorithms used for the prediction of the mid price direction. In Section~\ref{sec:evaluation} the experimental setup and results are described and analyzed. Finally, Section~\ref{sec:conclusion} summarizes the conclusions drawn from studying the application of the described Machine Learning algorithms to financial data.

\section{Related Work}
\label{sec:related_work}

The dynamics of the high frequency limit order book comprise a challenging field of study which has been investigated in past literature. %R3C1
An extensive survey on stochastic models and statistical techniques for modeling high frequency limit order book data can be found in \cite{cont2011statistical}, highlighting the inadequacies of statistical models as well as the need for more complex models, such as Machine Learning ones. Statistical models often make unrealistic assumptions about the distribution of the data, such as assuming a Poisson distribution of limit order events. Machine Learning techniques make no assumptions on the distribution of the data. Furthermore, the distribution of limit order events changes rapidly, not just from one day to the next but also within the same day. It is thus very challenging for statistical models which typically assume stationary signals to be used effectively used for modeling limit order book data. The drawbacks of statistical models and advantages of Machine Learning approaches have also been examined in \cite{kercheval2015modelling}. An extensive analysis of high frequency financial data can be found at \cite{engle2004analysis}, and the dynamics of limit order books are explained in detail in \cite{cont2011statistical,bouchaud2002statistical}.

Machine learning has been used very extensively to analyze the financial market from many different aspects. NNs and SVMs have been two of the most popular techniques for this task, as indicated by a number of works, e.g., \cite{kaastra1995forecasting,kaastra1996designing, tay2001application,cao2003support,lu2009financial}, which use models based on these architectures. In \cite{huang2005forecasting} an SVM model is trained to predict the direction of the movement of the NIKKEI 225 index. An SVM and Multilayer Perceptron (MLP) comparison can be found in \cite{kim2003financial}, where daily direction of the price of the Korea Composite stock index is predicted, using 12 different indexes as input features. Using two different windows, one long and one short term, in order to capture both the trend and higher frequency information of the time series of treasury bond returns, \cite{zhang2001multiresolution} utilizes an MLP model and attempts to predict the movement of the future bonds returns.  In \cite{tran2019data} a feedforward neural network the structure of which is determined in a data-driven manner is used for mid-price direction predictcon. In \cite{cao2003support} the authors compare the performance of SVMs, MLPs and Radial Basis Function (RBF) NNs in predicting price changes of future asset contracts. In \cite{tran2017tensor} a tensor-based regression model is used, which is further extended for tensor-based NN classification in \cite{tran2018temporal}. In \cite{walczak2001empirical} different dataset sizes are used to train a neural network model and it is shown that using too many samples that span too far into the past can degrade the prediction quality. The evaluation comparison is based on the profit that each model produces. In Deep Portfolio Theory \cite{heaton2016deep}, the authors use autoencoders to optimize the performance of a portfolio and beat the profit benchmarks such as the biotechnology IBB Index.

In a similar fashion to this work, \cite{kercheval2015modelling} uses several handcrafted time sensitive and insensitive features, extracted from the limit order book. These features include bid-ask spreads and mid prices, price differences, mean prices and volumes, along with derivatives of the price and volume, average and relative intensity indicators, totaling to 144 different features.  However, in \cite{kercheval2015modelling} the proposed methods are evaluated on a very small dataset that contains about 400,000 order book rows. In contrast, in our case a large-scale dataset that contains information for 10 days and 5 stocks is used, with the raw data being more than 4 million samples.
%R1C1

To the best of our knowledge, this is the first extensive study on using different Machine Learning techniques, including feature extraction methods and various classifiers, for a stock mid price prediction problem using on large scale high frequency limit order book data.  The scope of the conducted experiments and the significance of the observed results indicate that Machine Learning highly befits this task. The presented baseline results carve the path towards future research in this field that can potentially achieve even more accurate and significant predictions. To this end, we provide strong, yet easy to use and tune, machine learning baselines, as well as guidelines on how they should be applied on limit order book data. In contrast to other studies, such as \cite{tran2017tensor, tran2018temporal, tran2019data, tsantekidis2017using, tsantekidis2017forecasting, passalis2017time}, that employ advanced Deep Learning architectures, e.g., recurrent and convolutional neural networks, which require exhaustive hyper-parameter tuning, this paper focuses on simpler and easier to use traditional machine learning approaches. Having solid machine learning baselines is an important step towards developing more powerful Deep Learning models, since they provide a valuable tool for validating the more advanced models. To this end, we also conducted rigorous tests and statistical validation of the presented methods to demonstrate the predictive capabilities of the developed models.

\section{High Frequency Limit Order Book}
\label{sec:LOB}

In financial equity markets a limit order is a type of order to buy or sell a specific number of shares within a set price. For example, a sell limit order (ask) of $\$10$ with volume of 100 indicates that the seller wishes to sell the 100 shares for no less that $\$10$ a piece.  Respectively, a buy limit order (bid) of $\$10$ means that the buyer wishes to buy a specified amount of shares for no more than $\$10$ each. An in-depth survey of the properties of the LOB can be found in \cite{cont2011statistical}.

Consequently, the order book, which contains the above information, has two sides: the bid side, containing buy orders with prices~$\mathbf{p}_b(t)$ and volumes~$ \mathbf{v}_b(t)$, and the ask side, containing sell orders with prices~$\mathbf{p}_a(t)$ and volumes~$\mathbf{v}_a(t)$. The orders are sorted on both sides based on the price. On the bid side $p_b^{(1)}(t)$ is the highest available buy price and on the ask side $p_a^{(1)}(t)$ is the lowest available sell price.

Whenever a bid order price exceeds an ask order price $p_b^{(i)}(t) > p_a^{(j)}(t)$, where $p_b^{(i)}(t)$ is the $i$-th element of the bid side at time step $t$ and $p_a^{(j)}(t)$ is the $j$-th element of the ask side at the same time step, they ``annihilate", executing the orders and exchanging the traded assets between the investors. Typically, an order that leads to an immediate execution is called a market order. In this case, an investor makes an order to buy or sell a specific number of shares immediately, at the best available current price. Since the orders do not usually have the same requested volume, the order with the greater size remains in the order book with the remaining unfulfilled volume.

Several tasks arise from this data, ranging from the prediction of the price trend and the regression of the future value of a metric, e.g., volatility, to the detection of anomalous events that cause price jumps, either upwards or downwards. These tasks can lead to interesting applications, such as protecting the investments when market conditions are unreliable, or taking advantage of such conditions to create automated trading techniques for profit.

The data used in this work consists of $10$ orders for each side of the LOB. Each order is described by $2$ values, the price and the volume, yielding a total of $40$ values for each time step The stock data, provided by Nasdaq Nordic, come from the Finnish companies Kesko Oyj, Outokumpu Oyj, Sampo, Rautaruukki and W\"artsil\"a Oyj. The time period used for collecting that data ranges from the 1st to the 14th June 2010 (only business days are included), and the data is provided by the Nasdaq Nordic data feeds \cite{siikanen2017limit,siikanen2017drives}. The dataset is made up of $10$ days for $5$ different stocks and the total number of messages is about $4.5$ million with equally many separate depths. 

\section{Proposed Methodology}
\label{sec:proposed_methodology}
In this Section, we briefly review the data preprocessing and the feature extraction procedure. Then, we introduce two feature learning methods that are used to \textit{learn} low-dimensional features using the extracted handcrafted features. Finally, we introduce the classifications methods that are used to predict the mid price movements.

\subsection{Handcrafted Features}
\label{sec:handcrafted}

%R2C1

The raw order book data is first preprocessed by removing the unnecessary messages from the exchange, e.g., event messages, and then the features proposed in \cite{kercheval2015modelling} are extracted. More specifically, first, a basic set of features which includes the prices and volumes for every level of the ask and bid side of the order book is extracted. This information yields 40 values at each time step. Then, time-insensitive features describing the spread, mid-price, price and accumulated price differences between the bid and ask orders of each depth level, and  price and volume spreads are extracted. Finally, time-sensitive features are extracted corresponding to the average intensity for trades, orders, cancellations, deletion, execution of visible limit orders and execution of hidden limit orders. This set of features also includes the price and volume average values at each level of the LOB, the average intensity per trading type as well as comparisons between the intensities and limit activity acceleration (derivatives of average intensities). Because of the non-linear nature of time in the LOB data, we follow an event-based inflow. The interested reader is referred to \cite{kercheval2015modelling,ntakaris2017benchmark} for a more detailed description of the extracted features. Note that one feature vector is extracted for every 10 limit order events that change the LOB, effectively subsampling the data by a factor of 10. The total number of the collected limit order events is about 4.5 million, leading to a total of 453.975 extracted feature vectors.

\begin{figure}
	\centering
	\includegraphics[width=0.7\linewidth]{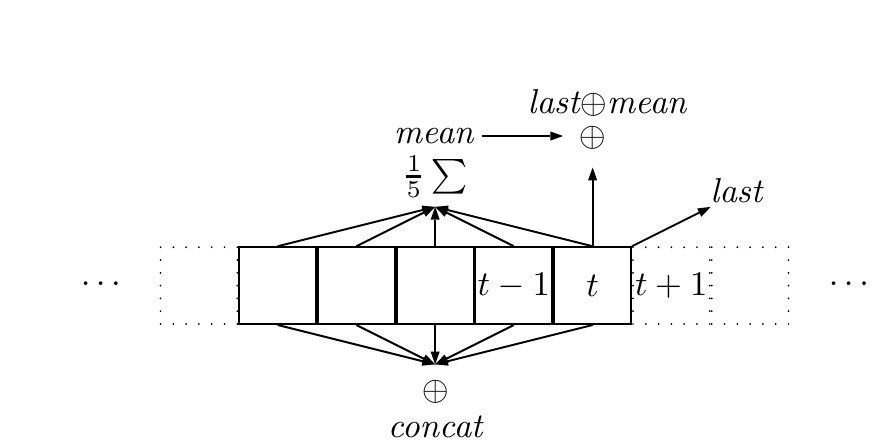}
	\caption{Representations extracted by using a sliding window of length 5 on the handcrafted feature vectors. The \emph{last$\oplus$mean} representation is the concatenation of the depicted \emph{last} and \emph{mean} representations.}
	\label{fig:representations}
\end{figure}

Instead of using only the feature vector extracted from the current time step, as proposed in \cite{kercheval2015modelling}, we propose three additional ways to extract representations capable of capturing more temporal information. Therefore, the following four different feature vectors are produced and used as inputs to the evaluated models, using a time sliding window of length $5$:
\begin{enumerate}
	\item A single feature vector with $144$ values as described above, corresponding to the last sample in the sliding window (abbreviated as \emph{last}).
	\item The mean of the $5$ samples currently in the window, which is also a $144$-dimensional vector (abbreviated as \emph{mean}).
	\item The concatenation of the last sample and the mean of all $5$ samples in the window, yielding a $288$-dimensional feature vector (abbreviated as  \emph{last$\oplus$mean} -- the circled plus symbol is used to denote the concatenation operation).
	\item The concatenation of all $5$ samples, yielding a $720$-dimensional feature vector (abbreviated as \emph{concat}).
\end{enumerate}
Figure \ref{fig:representations} illustrates the process of obtaining the above representations.

Using this sliding window approach allows for temporal information to be incorporated into the representations, which are subsequently used to predict the movement of the stock's mid price. By averaging over the five entries contained in the window, some of the inherent noise of the features is alleviated. Concatenating different representations, e.g., the last feature vector and the average of the last 5 feature vectors, allows for more temporal information to be introduced in the final feature vector.

Since normalization is crucial for most ML techniques, we normalize the extracted features using z-score standardization:

\begin{equation}
\mathbf{x} = \frac{\mathbf{x}_\text{raw} - \bar{\mathbf{x}}_\text{raw}}{\sigma_\text{raw}}
\end{equation}
where $\mathbf{x}_\text{raw}$ is the vector of values to be normalized, $\bar{\mathbf{x}}_\text{raw}$ is the mean vector of the data and $\sigma_\text{raw}$ is the standard deviation vector of the data. Both the mean and the standard deviation are computed element-wise.

\subsection{Prediction Labels}
\label{sec:targets}

Of the values accompanying LOB data, the tick price, which is the price of the last executed trade, typically varies wildly between the two sides of the margins, introducing great amounts of noise to the prediction labels. The so-called market micro-structure noise can be partially reduced by using mid-prices, i.e. the mean of the best ask and best bid prices, instead of transaction prices. Thus, in this paper, we considered mid-prices as the stock price observations. Another advantage of using mid-prices instead of transaction prices is that they are observable every time as long as there are orders on both bid and ask sides while transaction prices are updated only at transactions. The mid price $p(t)$ is defined as:
\begin{equation}
p(t) = \dfrac{p_a^{(1)}(t) + p_b^{(1)}(t)}{2}
\end{equation}
where $p_a^{(1)}$ and  $p_b^{(1)}$ are the best bid and best ask price. Note that the mid price of the stock is one of the features in the set of 144 features derived by following the feature crafting process described in \cite{kercheval2015modelling}.

An averaging filter is applied over the past $N_{\beta}$ values (including the current time step $t$) of the mid price of the samples to reduce the impact of the noise in the signal:

\begin{equation}
\label{eq:smoothed_mid_price}
m_{\beta}(t) = \dfrac{1}{N_{\beta}} \sum_{i=1}^{N_{\beta}} p(t-i+1)
\end{equation}
In our experiments we use $N_{\beta}=9$. We have also evaluated the models for different values of $N_{\beta}$, $5, 7$ and $11$, obtaining similar results.
%R3C2

We compare this price at time $t$ with the mean of the succeeding $N_{\alpha}$ smoothed mid prices (not including the current time step $t$):
\begin{equation}
\label{ref:succeeding_prices}
m_{\alpha}(t) = \dfrac{1}{N_{\alpha}} \sum_{i=1}^{N_{\alpha}} m_{\beta}(t+i)
\end{equation}

For each sample, the movement of the mid price is defined by comparing the current smoothed mid price $m_{\beta}(t)$ to the mean of the next $N_{\alpha}$ smoothed mid prices $m_{\alpha}(t)$. Small changes of the price should be considered insignificant. To this end, we introduce a parameter $\gamma$ to control the threshold a price movement must surpass in order to be considered as either upwards or downwards. Thus, the label $l(t)$ to be predicted for time step $t$ is computed as:

\begin{equation}
\label{eq:labels}
l(t) =
\begin{cases}
1\ \ ,\quad \text{if } m_{\alpha}(t) > m_{\beta}(t) \cdot (1 + \gamma) 
\quad \\
-1 ,\quad \text{if } m_{\alpha}(t) < m_{\beta}(t) \cdot (1 - \gamma) \\
0\ \  ,\quad \text{otherwise}
\end{cases}
\end{equation}

In other words, we treat the problem of forecasting the mid price movements of stocks as a classification problem with three possible outcomes: upwards movement, downwards movement and no change with labels $1, -1, $ and $0$ respectively as defined in Equation \eqref{eq:labels}. Three sets of such labels are generated through the above process, for $N_\alpha=1, 5 \text{ and } 10$ and $\gamma=0.0001, 0.0002 \text{ and } 0.0003$ respectively. All the classifiers were extensively evaluated using these three different prediction horizons.
%R3C2

In practice, the mid price of a stock very rarely remains the same through consecutive time steps, thus for $\gamma=0$ the majority of the mid price movements are categorized as either upwards of downwards, depleting the no-change class of samples. The value of the $\gamma$ parameter can be crucial as even small fluctuations significantly affect the balance between the three possible classes. As $\gamma$ increases, so does the number of samples belonging to the no-change class. Although raising this threshold would lead to a more balanced problem, i.e., all three classes would contain about the same number of samples, high values of $\gamma$ are undesirable as they allow upwards and downwards movements to be categorized as not having changed even though the movement could be significant. Choosing a meaningful value for $\gamma$ constitutes a trade-off between the balance among classes and the meaningfulness of the produced labels.

The values used for $\gamma$ are minuscule, which is meaningful since larger values would deprive the upwards and downwards classes of samples. On the other hand, smaller values would introduce a lot of noise in these classes by causing insignificant changes to be considered significant. Nonetheless, the selected values still produce an imbalanced dataset with most of the samples being classified as not having changed.

The resulting class imbalance constitutes a problem which must be taken into consideration by the ML algorithms used for the classification task. In this work, we deal with this problem by introducing class weights inversely proportional to the number of samples in each class. The interested reader is referred to \cite{galar2012review}, and \cite{longadge2013class}, for a detailed review of techniques that can be used to deal with the problem of class imbalance.

\subsection{Feature Learning}
\label{sec:feature_learning}
Apart from the four handcrafted feature vectors proposed in the previous subsection, two \textit{feature learning} methods were deployed and evaluated in our experiments. The methods and the details of their application to the problem at hand are described below. The extracted representations are used as inputs to the evaluated classifiers, on their own or in combination with the described representations created from the handcrafted features. 

\subsubsection{Autoencoders}

Autoencoders (AEs) are neural networks which map their input data to itself, through multiple levels of non-linear neurons \cite{vincent2008extracting, hinton2006reducing, ngiam2011multimodal}. Thus, the input and output layers consist of as many neurons as the dimension of the data. Such networks are comprised of an encoding part, which maps the input to an intermediate representation, and a decoding part, which maps the intermediate representation learned to the desired output and is symmetrical to the encoding part layer-wise.

Typically, an AE is used for dimensionality reduction as well as feature extraction, which means that the intermediate representation learned is lower-dimensional than the input data. The layers of both parts of the network $l=1,\dots,l_\text{enc},\dots,L$, where $l_\text{enc}$ is the encoding layer, are accompanied by weights $\mathbf{W}^{(l)}$ which multiply each layer's input to produce an output. A bias term $\mathbf{b}^{(l)}$ is also added to the output of the neuron, and a non-linearity $s(\cdot)$ called the activation function of the neuron is applied to this output to produce the neuron's activation value. The output $\mathbf{x}_\text{out}^{(l)}$ of the $l$-th layer is given by:
\begin{equation}
\mathbf{x}_\text{out}^{(l)} = s(\mathbf{W}^{(l)} \mathbf{x}_\text{in}^{(l)} + \mathbf{b}^{(l)})
\end{equation}
where $\mathbf{x}_\text{in}^{(l)}$ is the input to the $l$-th layer, which is equal to the output of the previous layer, or:
\begin{equation}
\label{eq:nn_input}
\mathbf{x}_\text{in}^{(l)} =
\begin{cases}
\mathbf{x}\quad \quad ,\quad l=0\\
\mathbf{x}_\text{out}^{(l-1)},\quad l>0\\
\end{cases}
\end{equation}
where $\mathbf{x} \in \mathbb{R}^D$ denotes an input sample. The \emph{last} representation is used as the input to the AE, i.e., the original input dimension is $D=144$.

The network's parameters can be learned using the well-known backpropagation algorithm \cite{rumelhart1988learning}, combined with an optimization method, such as Stochastic Gradient Descent (SGD) \cite{zhang2004solving}, with the final objective being the optimization of the network's loss function. Specifically for autoencoders, their objective is to minimize the reconstruction error, i.e., the squared $l_2$-norm between the network's output $\mathbf{x}_\text{out}^{(L)}$ and the desired output $\mathbf{x}$, which is the same as the network's input:

\begin{equation}
\ell = \|\mathbf{x}-\mathbf{x}_\text{out}^{(L)}\|_2^2
\end{equation}
The objective of the network is to minimize the mean of errors over all data samples.

As the training process converges, the activations of the intermediate layers can be used as learned feature representations of the input data. Let $\mathbf{x}_\text{enc}$ denote the output of the $l_\text{enc}$ layer:
\begin{equation}
\mathbf{x}_\text{enc}  = \mathbf{x}_\text{out}^{(l_\text{enc})} \in \mathbb{R}^d, d<D.
\end{equation}
Then $\mathbf{x}_\text{enc}$ can be used as the low-dimensional representation of the data in the subsequent classification task.

\subsubsection{Bag-of-Features}

Bag-of-Features (BoF) models \cite{sivic2003video, passalis2016information}, allow for extracting constant-length representations of samples that consist of multiple feature vectors, e.g., feature vectors extracted from various locations of an image or from various time points of a time series, such as the 5 feature vectors contained in the current sliding window. BoF models, originate from and comprise an extension of the standard Bag-of-Words model~\cite{salton1983mcgill}, which uses word frequencies as features to describe each document.  Similarly, the BoF model describes each sample as a histogram over a set of predefined codewords, which is also called \textit{codebook} or \textit{dictionary}.

To encode our data using the Bag-of-Features model, we must first learn the dictionary. To this end, we pick a random subsample of the data and apply the $k$-means clustering algorithm to find $K$ centers that best partition the data into clusters \cite{kanungo2002efficient}. The $k$-means  algorithm firstly picks $K$ random cluster centers and assigns each sample to the cluster whose center lies the closest to it. The cluster centers are then updated to be the mean of the samples belonging to each cluster and the process is repeated until the centers converge. The final cluster centers $\mathbf{v}_k, k=1,\dots,K$ form the dictionary of the BoF model.

The learned clusters act as histogram bins in which the feature vectors are quantized. For every sample to be encoded we compute its similarity to each of the codewords (cluster centers) as:
\begin{equation}
\label{eq:bof-1}
{d}_{i,k} = \exp \left(\frac{-\|\mathbf{v}_k -\mathbf{x}_{i}\|_2}{g} \right)
\end{equation}
where $\mathbf{x}_{i}$ is the $i$-th ($i=1,\dots,5$) feature vector of the current window, ${d}_{i,k}$ denotes the $k$-th element of the $\mathbf{d}_{i}$ vector and $g$ is a scaling parameter which controls the participation score of the feature vectors to each bin. All five samples contained in the sliding window as described in Section \ref{sec:handcrafted} are used for this process. Adjusting the scaling parameters $g$ alters the \textit{fuzziness} of the quantization process, with larger values leading to more fuzzy assignment \cite{nbof}. Equation (\ref{eq:bof-1}) is then normalized to have unit $l_1$ norm as follows:
\begin{equation}
\mathbf{u}_{i} = \frac{\mathbf{d}_{i}}{\| \mathbf{d}_{i} \|_1} \end{equation}
The vector $\mathbf{u}_{i}$ expresses the membership of the $i$-th feature vector to each of the clusters.

To obtain the final histogram representation $\mathbf{h}$ of the sample we average over the membership vectors $\mathbf{u}_{i}$:
\begin{equation}
\mathbf{h} = \frac{1}{N_F} \sum_{i=1}^{N_F} \mathbf{u}_{i} 
\end{equation}
where $N_F$ is the length of the current window, i.e., $N_F=5$ in this work. This histogram vector $\mathbf{h}$ can be then used for the subsequent classifications tasks alone or in combination with the other extracted feature vectors.

\subsection{Classifiers}
Three different classifiers were evaluated, combined with eight sensible combinations of the handcrafted input data and the features learned from that data.

\subsubsection{Support Vector Machines}
Support Vector Machines are binary, linear classifiers of the form \mbox{$f(\mathbf{x}) = \mathrm{sign}(\mathbf{w}^T\mathbf{x}+b)$} whose aim is to maximize the margin of the hyperplane separating the two classes. The vector $\mathbf{w} $ is orthogonal to the separating hyperplane, while the offset of the hyperplane is determined by the value of $b$. Multiple such binary classifiers can be used to solve multi-class classification problems. The optimization problem can be formulated as:

\begin{equation}
\label{eq:svm_criterion}
\begin{split}
\min_{\mathbf{w}\in\mathbb{R}^d, \xi_i\in\mathbb{R}^+} \|\mathbf{w}\|^2_2 + C\sum_{i=1}^n \xi_i \\
\text{s.t. } y_i (\mathbf{w}^T\mathbf{x}_i+b) \geq 1 - \xi_i, \forall i
\end{split}
\end{equation}
where $\xi_i, i=1,\dots,n$ are slack variables which allow the classifier to achieve better generalization, $C$ is a regularization parameter which controls the width of the margin learned by the SVM, and $y_i\in\{-1,1\}$ are the binary classification targets. In our experiments, the parameter $C$ was selected by performing 3-fold cross-validation on the training set (possible $C$ values range from $0.00001$ to $0.1$).
%R3C3

For a multiclass classification problem with unbalanced label distributions over the data, the multiple SVM classifiers trained can have different $C$ values. Choosing large values for the separation of less represented classes will lead to fewer misclassified samples. This is important to avoid deteriorated class-wise performance of the classifier for the less represented classes. To account for the class-imbalance in the training set, the regularizer is set to be inversely proportional to the frequency of each class in the training dataset, i.e., $C_i = \frac{1}{3}\frac{N}{N_{c_i}} C$, where $C_i$ is the regularizer for the SVM responsible of recognizing samples of the $i$-th class, $N$ is the total number of training samples, $N_{c_i}$ is the number of training samples that belong to the $i$-th class  and $C$ the global regularization parameter selected using cross-validation.
%R3C3

The direct solution of the above optimization problem requires the storage in memory of an $N\times N$ matrix containing the inner products between all pairs of samples, where $N$ is the number of samples in the training set. This can prohibit the computation of the direct solution when the dataset is large, as in our case. To combat this issue, gradient descent-based optimization techniques have been utilized, including SGD \cite{bousquet2008tradeoffs}. The learning process of following the gradient of the optimization objective might end near but not exactly at the global minimum. However, it allows for training the classifier by using minibatches of data \cite{hoi2009semisupervised}, thus leading to better generalization error by allowing for more samples to contribute towards the classifier's learning process.

\subsubsection{Single Hidden Layer Feedforward Neural Networks} Non-linear kernel methods, such as Kernel SVMs~\cite{hofmann2008kernel}, and Kernel Ridge Regression~\cite{robert2014machine}, can significantly increase the classification accuracy over their linear counterparts. However, these kernel methods are even more computationally intensive than their linear variants, requiring the calculation of the kernel matrix between all the training samples. To alleviate this problem, approximate methods have been proposed, such as Prototype Vector Machines~\cite{zhang2009prototype}, and Approximate Kernel Extreme Learning Machines~\cite{Iosifidis2017210}. These methods employ Single Hidden Layer Feedforward Neural Networks (SLFNs) to approximate the kernel solution through a non-linear hidden layer. 

In this work, a max-margin SLFN formulation is used,  i.e., the output layer is trained using a max-margin objective, as in \cite{iosifidis2015sparse}. First, the hidden layer weights are learned by clustering the data into $N_H$ clusters, where the centroid $\mathbf{w}_k$ of each cluster corresponds to a prototype vector. The activation of the $k$-th hidden neuron $x_{hid, k}$ is calculated by measuring the similarity between the input vector $\mathbf{x}$ to each prototype vector $\mathbf{w}_k$ using a Radial Basis Function (RBF):
\begin{equation}
x_{\text{hid}, k} = \exp \left(\frac{||\mathbf{x} - \mathbf{w}_k||^2_2}{2 \sigma^2} \right)
\end{equation}
where $\sigma$ is scaling parameter that alters the spread of RBFs. Typically $\sigma$ is set to the mean distance between the input samples. Then, the output of each binary classifier is calculated as:
\begin{equation}
f(\mathbf{x}) = \mathbf{W}\mathbf{x}_\text{hid}+\mathbf{b}
\end{equation}
where $\mathbf{W}$ and  $\mathbf{b}$ are the weights and the biases of the output layer. The output weights are learned by using gradient descent and a max-margin objective (as described in Equation (\ref{eq:svm_criterion})). As before, the regularization parameter $C$ is selected using 3-fold cross-validation and appropriately weighted considering the distribution of the class labels.
%R3C3

\subsubsection{Multilayer Perceptrons}
Multilayer Perceptrons (MLPs) \cite{haykin2009neural}, also known as Multilayer Feedforward Neural Networks, consist of several layers of weighted connections through which the input is processed. A non-linear activation function is applied after each layer. The calculation of each layer's output is similar to the autoencoder's:

\begin{equation}
\label{eq:mlp_forward}
\mathbf{x}_\text{out}^{(l)} = s(\mathbf{W}^{(l)}\mathbf{x}^{(l)}_\text{in} + \mathbf{b}^{(l)})
\end{equation}
where $\mathbf{x}^{(l)}_\text{in}$ is the input to a layer, obtained by Equation~(\ref{eq:nn_input}), $s(\cdot)$ is the activation function, $\mathbf{W}^{(l)}$ is the weight matrix accompanying the current layer $l$, and $\mathbf{b}^{(l)}$ is the current layer's bias vector. Equation (\ref{eq:mlp_forward}) is applied for every layer in the neural network up until the last layer $l_\text{out}$ where each neuron represents a different class, meaning that in the $l_\text{out}$ there must be as many neurons as classes in our dataset, i.e., $N_C=3$ neurons. The softmax function $\sigma(\cdot)$ is applied to the output of the final layer of the MLP, to produce a probability distribution over the existing classes:
\begin{equation}
\sigma(\mathbf{x}_\text{out})_j = \frac{e^{x_{\text{out},j}}}{\sum_{k=1}^{N_C} e^{x_{\text{out}, k}}}
\end{equation}
where $\sigma(\mathbf{x}_\text{out})_j$ is the predicted probability for class $j$ and $x_{out,j}$ is the output of the $j$-th neuron of the network.

To encode the prediction objective as a differentiable cost function, the categorical cross entropy function is used:
\begin{equation}
\label{eq:nll}
\ell = -\sum_j y_j \log \sigma(\mathbf{x}_\text{out})_j
\end{equation}
where $y_j$ is the desired output for the $j$-th output neuron,  $\mathbf{x}_\text{out}$ denotes the output of the MLP, and $\sigma(\mathbf{x}_{\text{out}})_j$ is the predicted probability for the $j$-th class. As with AEs, the weights of each layer can be learned using SGD. However, instead of using plain SGD to optimize the parameters of the MLP, the ADAM algorithm~\cite{kingma2014adam}, which is a more advanced optimization algorithm that allows for faster and more stable convergence, is used.

\section{Experimental Evaluation}
\label{sec:evaluation}

\subsection{Evaluation Setup}

\begin{figure*}
	\begin{center}
		\includegraphics[width=.85\linewidth]{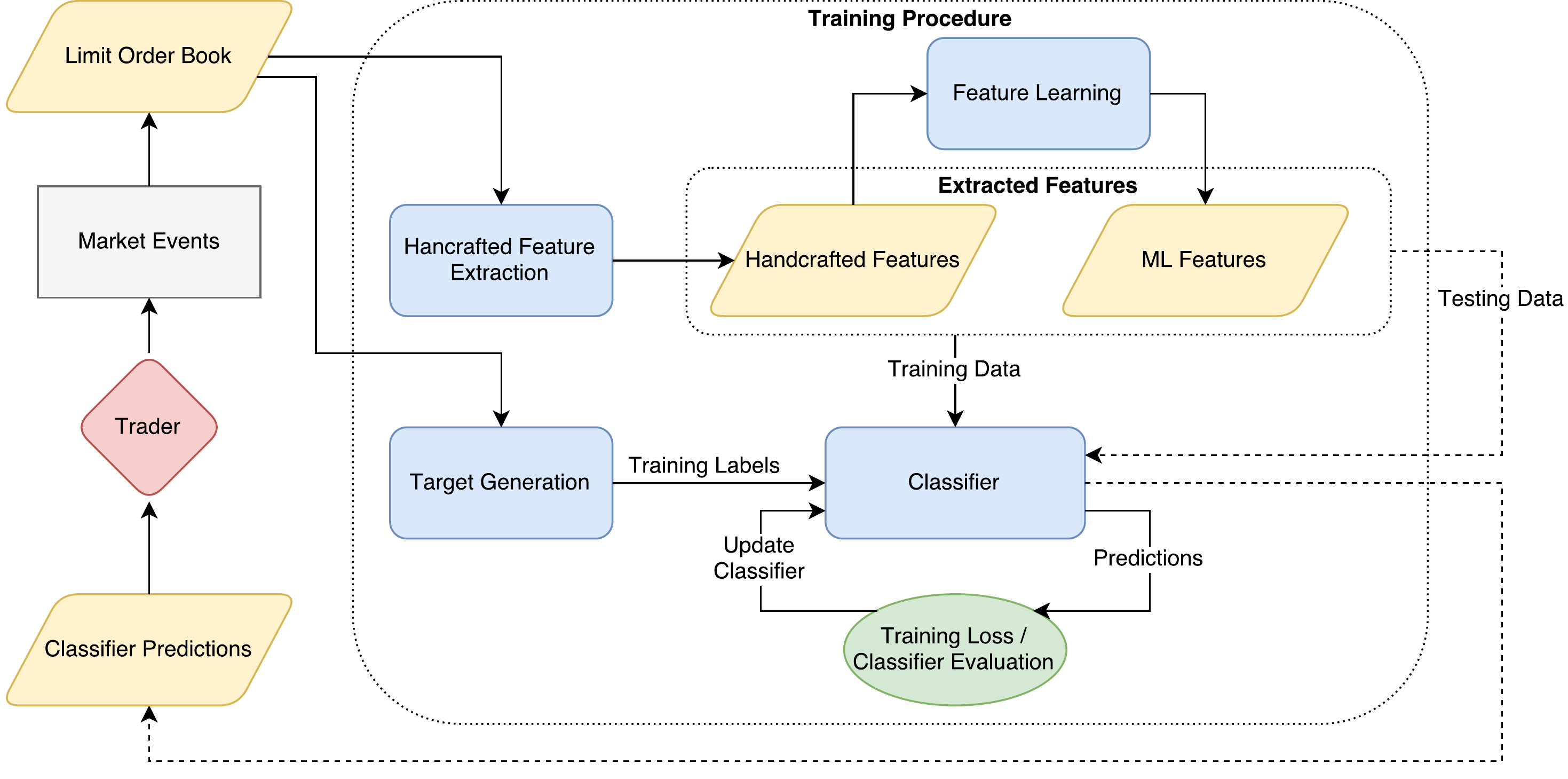}
		\caption{Pipeline of the stock mid price prediction system used. Limit order events generated by market events are used to extract feature vectors which describe the stocks present in the LOB, and a classifier is trained to recognize the movement of the mid-price of those stocks. Dashed lines indicate the deployment phase, where a trader receives classification predictions and potentially acts on them thus influencing the market.}
		\label{fig:pipeline}
	\end{center}	
\end{figure*}

%R2C5
An overview of our proposed system of analysis is shown in Figure~\ref{fig:pipeline}. Events taking place in the market are described in the limit order book, from which we can then extract handcrafted features as described in Section~\ref{sec:handcrafted} as well as target values corresponding to the movement of the mid-price. The data used consist of combinations of the handcrafted feature vectors and features learned from these as discussed in Section~\ref{sec:feature_learning}. The classifiers are then trained incrementally using minibatches of data. Once training has converged, the classifier makes predictions about unseen data which may influence the behavior of a trader, who then takes place in creating market events. To evaluate the performance of our models two different experimental setups are used, which are described below.

\subsubsection{Anchored walk forward}
\begin{figure}[ht]
	\centering
	\includegraphics[width=0.5\linewidth]{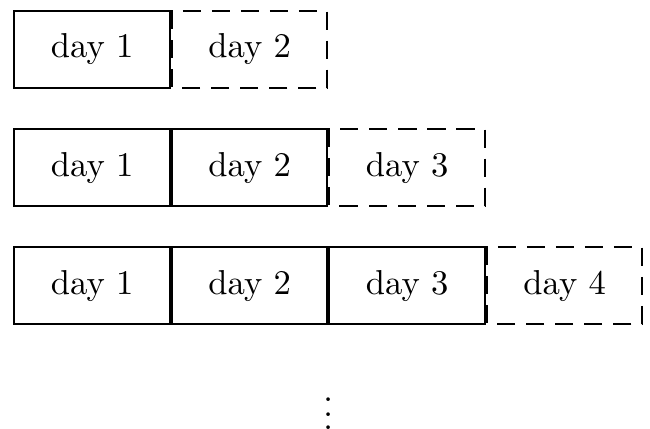}
	\caption{In the anchored walk forward evaluation method, the first $d=1,\dots 9$ days are used progressively as the training dataset and evaluation is performed on the $(d+1)$-th day. The solid line is used to depict days used for the training dataset, whereas the dashed line indicates the day used for the evaluation. The first three progressions of this evaluation setup are shown.}
	\label{fig:anchored_walk_forward}
\end{figure}

In this case, the first $d=~1,\dots 9$ day samples of all the stocks are progressively used to train our models and the sample belonging to the $(d+1)$-th day is used for testing. Since our data consists of 10 days for each stock, this means we have a total of 9 different folds to run our models. This evaluation method is known as \emph{anchored walk forward analysis}, as the starting point in time remains fixed \cite{tomasini2011trading}. Figure \ref{fig:anchored_walk_forward} illustrates the first three progressions of this evaluation method. By using this evaluation method, we can determine whether the models are able to learn features which capture temporal information, so as to be able to predict future mid price movements.

\subsubsection{Hold-out per stock}

For this setup, the models are trained on data describing 4 stocks and evaluated on the last unknown stock. The experiments are executed 5 times so that every stock is used as the unseen stock the models are evaluated on. This evaluation method is used to determine whether the models can learn features from stocks that can be applied to an unknown stock.

\subsection{Evaluation Results}

Three metrics are used for the evaluation of the described models: \emph{average precision per class} (macro precision), \emph{average recall per class} (macro recall) and \emph{average F-score} (macro F-score). The precision is defined as the ratio of the true positives over the sum of the true positives and the false positives, while the recall is the ratio of the true positives over the sum of the true positives and the false negatives. The F-score is defined as the harmonic mean of the precision and the recall. However, as the F-score is also macro-averaged over the three classes, its values might lie outside the range of the mean of the precision and recall.

Because of the class imbalance caused by considering only meaningful changes in the movement of the mid prices, a classifier biased towards the no-change class would achieve very high accuracy, as the majority of the predicted labels would match the ground truth. The accuracy metric becomes meaningless in such severely imbalanced problems and is omitted. For the macro-averaged metrics, the corresponding metric is first computed for each class and finally all three values --- one for each class --- are averaged. Thus, low results in the less represented classes will equally affect the final result.

For the AE representation, a $5$-layer architecture of $144$-$72$-$24$-$72$-$144$ neurons for each respective layer is used. The intermediate $24$-dimensional representation $\mathbf{x}_\text{enc}$ is extracted and used as input to the evaluated classifiers. As for the BoF model, the fuzziness parameter $g$ is set to $0.01$ and $K$ is equal to $128$, thus producing $128$-dimensional histograms to be used as representations of the input for the classification task.

The representations obtained by the sliding window over the handcrafted features in combination with the features extracted by the AE, and BoF models are used as the input to the described classifiers. The circled plus symbol is used to denote the concatenation operation in Tables \ref{tab:svm_day_results}-\ref{tab:mlp_stock_results} which summarize the evaluation results. For example, \emph{last$\oplus$BoF} denotes the concatenation of the \emph{last} representation with the representation extracted by the BoF model. To evaluate the models under a wide range of conditions we have conducted extensive experiments using three different prediction horizons ($N_a$), i.e., for $N_\alpha=1$, $N_\alpha=5$ and $N_\alpha=10$, and for both of the experimental setups discussed.
%R3C2

\subsubsection{SVM Results}

For the linear SVM classifier trained with SGD, the class imbalance accompanying the data is rectified by introducing weights associated with each class and adjusting them to be inversely proportional to the number of samples belonging to that class. This ensures that the less represented classes will be taken into account in the optimization process, making the classifier less biased towards the better represented class and thus more useful for practical applications.

Table~\ref{tab:svm_day_results} contains the results achieved for all three different sets of prediction targets for the anchored day evaluation setup, whereas Table~\ref{tab:svm_stock_results} contains the results for stock hold-out setup. The $N_\alpha$ column contains the number of succeeding samples taken into consideration for the production of the prediction targets, as discussed in Section \ref{sec:targets}. The \emph{Input} column contains the representation used as input to the classifier. The macro-averaged precision, recall and F-score are presented, as well as the standard deviation observed for these metrics over the progressive experiments for both evaluation setups.

\begin{table}[hb!]
	\footnotesize
	\caption{Anchored walk forward evaluation: SVM}
	\label{tab:svm_day_results}
	\begin{center}
		\begin{tabular}{rl|ccc}
			$\mathbf{N_\alpha}$ & \normalfont{\textbf{Input}}  &  \textbf{Precision} &  \textbf{Recall } & \textbf{F-score}  \\
			\hline
			1& last  &  $45.72 \pm 5.83$ & $37.96 \pm 2.99$ & $37.51 \pm 2.45 $ \\
			1& mean  &  $48.14 \pm 5.08$ & $40.81 \pm 2.85$ & $41.53 \pm 1.78 $ \\
			1& last$\oplus$mean   &  $\mathbf{51.97 \pm 5.02}$ & $42.77 \pm 3.39$ & $43.95 \pm 1.68 $ \\
			1& concat &  $49.31 \pm 4.13$ & $\mathbf{44.65 \pm 2.32}$ & $\mathbf{45.44 \pm 1.87}$ \\
			\hline
			1 & AE  &  $40.17 \pm 7.21$ & $35.88 \pm 2.35$ & $34.32 \pm 2.06 $ \\
			
			1 & BoF  &  $44.50 \pm 3.45$ & $37.10 \pm 1.07$ & $37.73 \pm 1.21 $ \\
			1 &  AE$\oplus$BoF  &  $\mathbf{45.98 \pm 3.97}$ & $39.57 \pm 3.16$ & $38.16 \pm 2.99 $ \\
			1& last$\oplus$BoF  &  $45.32 \pm 4.17$ & $\mathbf{40.32 \pm 3.63}$ & $\mathbf{39.80 \pm 2.68}$ \\
			\hline
			\hline
			5 &  last  &  $40.47 \pm 1.88$ & $41.00 \pm 1.92$ & $38.56 \pm 1.72 $ \\
			5 &  mean  &  $50.60 \pm 3.29$ & $46.22 \pm 1.32$ & $46.91 \pm 1.72 $ \\
			5 &  last$\oplus$mean  &  $\mathbf{52.14 \pm 2.56}$ & $50.30 \pm 1.54$ & $\mathbf{50.68 \pm 1.39}$ \\
			5 &  concat  &  $50.95 \pm 3.83$ & $\mathbf{50.64 \pm 1.90}$ & $49.81 \pm 2.51 $ \\
			\hline
			5 &  AE  &  $42.66 \pm 4.04$ & $38.49 \pm 3.07$ & $36.75 \pm 2.58 $ \\
			5 &  BoF  &  $44.85 \pm 2.93$ & $41.73 \pm 1.14$ & $42.26 \pm 1.34 $ \\
			5 &  AE$\oplus$BoF  &  $\mathbf{47.64 \pm 1.96}$ & $42.57 \pm 1.57$ & $\mathbf{43.27 \pm 1.53}$ \\
			
			5 &  last$\oplus$BoF  &  $45.13 \pm 3.03$ & $\mathbf{44.16 \pm 2.35}$ & $42.43 \pm 1.78 $ \\
			\hline
			\hline
			10 &  last  &  $42.10 \pm 2.88$ & $40.86 \pm 1.80$ & $40.48 \pm 1.94 $ \\
			10 &  mean  &  $48.03 \pm 2.96$ & $47.28 \pm 2.48$ & $46.62 \pm 3.31 $ \\
			10 &  last$\oplus$mean &  $\mathbf{50.07 \pm 1.89}$ & $49.51 \pm 1.60$ & $\mathbf{49.25 \pm 1.58}$ \\
			10 &  concat  &  $49.74 \pm 3.15$ & $\mathbf{49.86 \pm 1.47}$ & $49.20 \pm 2.27 $ \\
			\hline
			10 &  AE  &  $39.83 \pm 2.72$ & $38.49 \pm 2.31$ & $36.01 \pm 1.84 $ \\
			
			10 &  BoF  &  $42.89 \pm 1.43$ & $42.01 \pm 1.82$ & $41.57 \pm 1.31 $ \\
			10 & AE$\oplus$BoF  &  $\mathbf{46.18 \pm 2.05}$ & $42.36 \pm 1.98$ & $41.70 \pm 2.45 $ \\
			
			10 & last$\oplus$BoF  &  $44.35 \pm 2.62$ & $\mathbf{44.27 \pm 1.91}$ & $\mathbf{42.70 \pm 1.39}$ \\
			\hline
		\end{tabular}
	\end{center}
\end{table}

\begin{table}[ht!]
	\footnotesize
	\caption{Hold-out per stock evaluation: SVM}
	\label{tab:svm_stock_results}
	\begin{center}
		\begin{tabular}{rl|ccc}
			$\mathbf{N_\alpha}$ & \normalfont{\textbf{Input}}  &  \textbf{Precision} &  \textbf{Recall } & \textbf{F-score}  \\
			\hline
			1 &  last  &  $41.41 \pm 5.48$ & $38.28 \pm 3.12$ & $31.03 \pm 9.08 $ \\
			1 &  mean  &  $46.24 \pm 3.14$ & $40.06 \pm 3.36$ & $40.02 \pm 2.63 $ \\
			1 &  last$\oplus$mean &  $48.04 \pm 3.67$ & $42.74 \pm 4.53$ & $\mathbf{41.49 \pm 2.69}$ \\
			1 &  concat  &  $\mathbf{48.98 \pm 6.28}$ & $\mathbf{44.19 \pm 7.81}$ & $38.85 \pm 6.36 $ \\
			\hline
			1 &  AE  &  $37.05 \pm 6.75$ & $34.78 \pm 1.70$ & $27.98 \pm 13.10 $ \\
			
			1 &  BoF  &  $40.78 \pm 3.94$ & $\mathbf{37.50 \pm 1.73}$ & $31.97 \pm 10.18 $ \\
			1 &  AE$\oplus$BoF  &  $41.63 \pm 2.37$ & $36.64 \pm 2.86$ & $34.85 \pm 2.74 $ \\
			
			1 &  last$\oplus$BoF  &  $\mathbf{42.97 \pm 3.41}$ & $37.26 \pm 2.18$ & $\mathbf{36.48 \pm 2.93}$ \\
			\hline
			\hline
			5 &  last  &  $40.58 \pm 2.65$ & $39.45 \pm 2.38$ & $36.08 \pm 5.50 $ \\
			5 &  mean  &  $47.37 \pm 2.61$ & $46.11 \pm 2.58$ & $45.00 \pm 2.12 $ \\
			5 &  last$\oplus$mean  &  $\mathbf{49.19 \pm 3.48}$ & $49.18 \pm 2.39$ & $47.59 \pm 1.72 $ \\
			5 &  concat  &  ${48.95 \pm 2.64}$ & $\mathbf{50.17 \pm 2.50}$ & $\mathbf{48.58 \pm 1.57}$ \\
			\hline
			5 &  AE  &  $38.93 \pm 1.43$ & $36.22 \pm 2.52$ & $27.78 \pm 11.62 $ \\

			5 &  BoF  &  $44.33 \pm 4.28$ & $40.61 \pm 3.89$ & $37.93 \pm 5.22 $ \\
			5 &  AE$\oplus$BoF  &  $43.02 \pm 2.01$ & $38.14 \pm 2.21$ & $37.77 \pm 3.00 $ \\

			5 &  last$\oplus$BoF  &  $\mathbf{44.76 \pm 2.86}$ & $\mathbf{40.85 \pm 2.62}$ & $\mathbf{40.20 \pm 2.80}$ \\
			\hline
			\hline
			10 &  last  &  $43.32 \pm 4.13$ & $38.66 \pm 1.96$ & $37.66 \pm 2.27 $ \\
			10 &  mean  &  $46.01 \pm 1.59$ & $46.69 \pm 3.84$ & $45.70 \pm 2.61 $ \\
			10 &  last$\oplus$mean  &  $\mathbf{48.89 \pm 2.40}$ & $\mathbf{47.91 \pm 2.40}$ & $\mathbf{47.74 \pm 1.90}$ \\
			10 &  concat  &  ${47.71 \pm 0.52}$ & $47.88 \pm 4.46$ & $45.14 \pm 4.58 $ \\
			\hline
			10 &  AE  &  $42.78 \pm 6.13$ & $35.38 \pm 1.17$ & $30.91 \pm 2.99 $ \\

			10 &  BoF  &  $40.48 \pm 1.44$ & $38.51 \pm 3.20$ & $33.34 \pm 9.69 $ \\
			10 &  AE$\oplus$BoF  &  $44.15 \pm 2.57$ & $37.38 \pm 3.71$ & $33.22 \pm 3.82 $ \\
			
			10 &  last$\oplus$BoF  &  $\mathbf{44.99 \pm 3.40}$ & $\mathbf{40.27 \pm 3.11}$ & $\mathbf{40.05 \pm 2.81}$ \\
			\hline\end{tabular}
	\end{center}
\end{table}

For the anchored day setup, the results indicate that information from the past and the present can be used to generalize and make useful predictions about future movements of the stocks' mid prices. Taking into consideration all of the metrics, the results seem to improve when the prediction target is derived as the mean of next $5$ and $10$ samples, and slightly deteriorate when the prediction target is the movement of the next $1$ sample. This means that the classifier is able to better capture the average movement of the mid price over a few succeeding time steps, which is expected as the movement of the mid price in the directly succeeding time step can be more noisy.

Moreover, for all prediction targets, as past values are taken into consideration, i.e., for all representations excluding \emph{last} and \emph{AE} but including their concatenations with the rest, the classifier's performance improves. Thus, information from not only the current time step but also from a few steps back seems to be important in the generalization of the classifier and the correctness of the predicted movements.

Reducing the dimensionality of the data, i.e., using the \emph{AE}, or the \emph{BoF} representation, improves the speed of the classification process but at the same time harms the prediction metrics. However, combining the learned features with the handcrafted features seems to improve the classification metrics. For example, combining the \emph{last} or the \emph{AE} features, that only contain information of the current time sample, with the \emph{BoF} representation, that captures the temporal dynamics of the stock as expressed in the last 5 time steps, improves the performance over simply using the \emph{last} features. This can provide useful insight for further developing deep learning techniques that will capture both the current tendency as well as the temporal progression of the time series data.

Furthermore, note that the performance of the SVM is further improved upon when the dimensionality of the data increases. This is reflected by the deteriorated performance of the \emph{AE} representation, where the input is only $24$-dimensional, as well as by the improved performance of the concatenated representations, e.g., \emph{last$\oplus$mean}, \emph{concat}, \emph{AE$\oplus$BoF}, where the dimensionality of the data is augmented by the concatenation operation. This behavior is to be expected when using a linear classifier, as data lying in high-dimensional spaces are more easily separated by linear hyperplanes.

As for the hold-out evaluation setup, the results demonstrate that information from other stocks can be utilized to make predictions for an unknown stock. Thus, the classifier captures the general trend of the stock-market by extracting information from a few stocks and is able to generalize to another stock, which hasn't contributed to its training process. As with the previous evaluation setup, the results seem to slightly improve when $N_\alpha$ is equal to $5$ or $10$, instead of $1$, reaffirming the stipulation that the average movement of the mid price over the next few samples is easier to predict than its direct movement in the next $1$ sample. 
However, the results for $N_a = 1$ are still significant and outperform a biased classifier. Note that a perfectly biased random classifier cannot consistently achieve a (macro-averaged) F-score higher than 33.33\% on the considered 3-class problems.
%R1C4

Once again, due to the linearity of the classifier, the lower-dimensional representations perform slightly more poorly than the higher-dimensional ones. The performance of the classifier is better when predicting the average movement of the mid price for the next $10$ samples and when using higher-dimensional representations, such as the \emph{concat} representation. The predictions made by the classifier reflect the general trend of the market rather than the individual tendencies of each stock. This is further corroborated by the high standard deviations in the F-score especially in the case where $N_\alpha=1$, which indicate that information only from other stocks is insufficient when making predictions about an unseen stock.

\subsubsection{SLFN Results}

For the SLFN, the size of the hidden layer is set to $1000$ and $\sigma$ is set to the mean pairwise distance between the training feature vectors. The $k$-means clustering algorithm is used for the computation of the weights of the hidden layer, whose activations are RBFs which measure the similarity between the input and each of the $1000$ prototype vectors learned. Similarly to the linear SVM, class weights are used to manage the imbalance between classes. The performance of this classifier is summarized in Table~\ref{tab:slfn_day_results} for the anchored day setup and in Table~\ref{tab:slfn_stock_results} for the hold-out setup.

For the anchored day setup, the performance of the classifier is more or less consistent for all three prediction targets. This means that the classifier is able to make generalizations about the movement of the mid price in the immediately succeeding time step, just as well as for its average movement in the next $5$ and $10$ time steps. In comparison to the  linear SVM classifier, the SLFN achieves somewhat better precision at the cost of slightly worse recall and F-score values. In other words, the predictions made by this model are more precise, but at the same time the model falsely classifies many positive samples as negative.

Although the model takes into consideration the class imbalance in the final classification task, it relies heavily on its first, unsupervised part which performs clustering on the input data without taking into account the distribution of the classes. Thus, the prototype vectors might be unfairly distributed over the classes, favoring one over the others, depending on the geometric distribution of the data in the input space. If an input sample is highly similar to one or more prototype vectors which the model has been trained to map to one of the classes, the model confidently classifies this sample as positive, i.e., as belonging to that class. Thus, such samples are more likely to be correctly classified, leading to more precise predictions. However, due to the unsupervised nature of the clustering step, an input sample is more likely to resemble prototype vectors mapped to multiple classes. Such samples lead the model to false negative predictions, accounting for the low recall scores. This phenomenon is even more severe when distribution shift and concept drift issues exist, as in the case of the hold-out evaluation setup (Table \ref{tab:slfn_stock_results}).

\begin{table}[ht!]
	\footnotesize
	\caption{Anchored walk forward evaluation: SLFN }
	\label{tab:slfn_day_results}
	\begin{center}
		\begin{tabular}{rl|ccc}
			$\mathbf{N_\alpha}$ & \normalfont{\textbf{Input}}  &  \textbf{Precision} &  \textbf{Recall } & \textbf{F-score}  \\
			\hline
			1 &  last &  $46.57 \pm 2.47$ & $41.19 \pm 4.00$ & $35.60 \pm 6.13 $ \\
			1 &  mean &  $48.74 \pm 3.74$ & $42.12 \pm 4.34$ & $36.67 \pm 11.06 $ \\
			1 &  last$\oplus$mean &  $\mathbf{50.46 \pm 3.59}$ & $\mathbf{37.52 \pm 1.35}$ & $\mathbf{38.78 \pm 2.04}$ \\
			1 &  concat &  $47.37 \pm 4.51$ & $35.44 \pm 1.18$ & $35.42 \pm 1.83 $ \\
			\hline
			1 &  AE &  $37.08 \pm 8.70$ & $\mathbf{36.22 \pm 3.49}$ & $31.57 \pm 4.16 $ \\
			
			1 &  BoF &  $34.20 \pm 9.12$ & $34.72 \pm 3.00$ & $33.02 \pm 2.93 $ \\
			1 &  AE$\oplus$BoF &  ${36.91 \pm 8.04}$ & $35.33 \pm 3.72$ & $\mathbf{33.03 \pm 2.24}$ \\

			1 &  last$\oplus$BoF &  $\mathbf{41.54 \pm 7.87}$ & $35.71 \pm 2.60$ & $32.60 \pm 7.97 $ \\
			\hline\hline
			5 &  last &  $48.80 \pm 2.76$ & $41.36 \pm 3.80$ & $34.74 \pm 6.61 $ \\
			5 &  mean &  $51.14 \pm 3.40$ & $\mathbf{46.57 \pm 3.61}$& $\mathbf{43.43 \pm 3.26}$ \\
			5 &  last$\oplus$mean &  $50.71 \pm 2.09$ & $43.95 \pm 4.45$ & $40.97 \pm 3.30 $ \\
			5 &  concat &  $\mathbf{51.29 \pm 4.27}$ & $36.49 \pm 2.03$ & $34.88 \pm 2.17 $ \\
			\hline
			5 &  AE &  $38.26 \pm 4.18$ & $37.08 \pm 3.60$ & $31.80 \pm 9.24 $ \\
			
			5 &  BoF &  $45.77 \pm 4.35$ & $41.39 \pm 4.89$ & $35.18 \pm 10.51 $ \\
			5 &  AE$\oplus$BoF &  $47.35 \pm 4.84$ & $37.45 \pm 4.28$ & $31.79 \pm 3.96 $ \\
			
			5 &  last$\oplus$BoF &  $\mathbf{49.49 \pm 2.15}$ & $\mathbf{41.66 \pm 5.69}$ & $\mathbf{35.29 \pm 7.72}$ \\
			\hline \hline
			10 &  last &  $48.01 \pm 3.81$ & $41.24 \pm 3.79$ & $35.34 \pm 7.08 $ \\
			10 &  mean &  $51.10 \pm 2.81$ & $\mathbf{44.66 \pm 5.87}$ & $\mathbf{38.56 \pm 11.27}$ \\
			10 &  last$\oplus$mean &  $51.59 \pm 2.69$ & $40.88 \pm 3.43$ & $37.97 \pm 9.87 $ \\
			10 &  concat &  $\mathbf{51.77 \pm 3.71}$ & $39.78 \pm 5.16$ & $37.75 \pm 6.19 $ \\
			\hline
			10 &  AE &  $39.61 \pm 12.32$ & $37.74 \pm 3.55$ & $31.44 \pm 5.08 $ \\
			
			10 &  BoF &  $47.41 \pm 3.03$ & $40.53 \pm 3.89$ & $35.05 \pm 7.32 $ \\
			10 &  AE$\oplus$BoF &  $47.26 \pm 2.76$ & $\mathbf{41.75 \pm 3.53}$ & $\mathbf{36.73 \pm 4.96}$ \\
			
			10 &  last$\oplus$BoF &  $\mathbf{48.46 \pm 3.70}$ & $41.25 \pm 5.34$ & $36.03 \pm 7.47 $ \\
			\hline
		\end{tabular}
	\end{center}
\end{table}

\begin{table}[ht!]
	\footnotesize
	\caption{Hold-out per stock evaluation: SLFN }
	\label{tab:slfn_stock_results}
	\begin{center}
		\begin{tabular}{rl|ccc}
			$\mathbf{N_\alpha}$ & \normalfont{\textbf{Input}}  &  \textbf{Precision} &  \textbf{Recall } & \textbf{F-score}  \\
			\hline
			1 &  last &  $42.76 \pm 3.99$ & $38.25 \pm 1.80$ & $28.09 \pm 13.06 $ \\
			1 &  mean &  $\mathbf{46.11 \pm 6.52}$ & $\mathbf{41.72 \pm 6.28}$ & $\mathbf{31.52 \pm 13.97}$ \\
			1 &  last$\oplus$mean &  $37.06 \pm 16.73$ & $40.53 \pm 5.95$ & $31.36 \pm 13.04 $ \\
			1 &  concat &  $45.58 \pm 6.84$ & $39.10 \pm 8.21$ & $30.93 \pm 6.34 $ \\
			\hline
			1 &  AE  &  $29.21 \pm 14.48$ & $34.58 \pm 1.58$ & $19.50 \pm 12.03 $ \\
			
			1 &  BoF &  $\mathbf{38.13 \pm 7.08}$ & $35.14 \pm 1.72$ & $28.10 \pm 13.60 $ \\
			1 &  AE$\oplus$BoF &  $36.31 \pm 5.49$ & $\mathbf{36.76 \pm 4.36}$ & $\mathbf{28.34 \pm 10.30}$ \\
			
			1 &  last$\oplus$BoF &  $38.07 \pm 5.62$ & $36.40 \pm 4.01$ & $27.21 \pm 11.19 $ \\
			
			\hline \hline
			5 &  last &  $44.54 \pm 4.12$ & $39.96 \pm 3.77$ & $31.62 \pm 8.66 $ \\
			5 &  mean &  $\mathbf{49.84 \pm 0.91}$ & $\mathbf{42.63 \pm 2.58}$ & $\mathbf{42.46 \pm 2.42}$ \\
			5 &  last$\oplus$mean &  $47.01 \pm 2.49$ & $39.31 \pm 5.28$ & $34.48 \pm 3.03 $ \\
			5 &  concat &  $44.52 \pm 9.78$ & $37.58 \pm 5.03$ & $28.38 \pm 7.92 $ \\
			\hline
			5 &  AE &  $34.82 \pm 8.76$ & $34.03 \pm 1.88$ & $18.91 \pm 8.42 $ \\
			
			5 &  BoF &  $42.82 \pm 3.34$ & $\mathbf{36.11 \pm 2.91}$ & ${30.28 \pm 4.13}$ \\
			5 &  AE$\oplus$BoF &  $\mathbf{43.26 \pm 3.64}$ & $35.26 \pm 1.83$ & $27.83 \pm 10.01 $ \\
			
			5 &  last$\oplus$BoF &  $42.15 \pm 6.46$ & $35.74 \pm 1.63$ & $\mathbf{33.40 \pm 4.21}$ \\
			\hline \hline
			10 &  last &  $43.99 \pm 4.85$ & $38.51 \pm 4.16$ & $22.80 \pm 11.72 $ \\
			10 &  mean &  $45.63 \pm 3.29$ & $39.48 \pm 4.49$ & $24.90 \pm 13.95 $ \\
			10 &  last$\oplus$mean &  $\mathbf{46.08 \pm 2.73}$ & $\mathbf{44.39 \pm 7.24}$ & $\mathbf{32.98 \pm 11.56}$ \\
			10 &  concat &  $35.25 \pm 15.57$ & $40.71 \pm 4.82$ & $29.95 \pm 10.68 $ \\
			\hline
			10 &  AE &  $31.52 \pm 9.10$ & $34.60 \pm 1.14$ & $22.65 \pm 7.20 $ \\
			
			10 &  BoF &  $42.10 \pm 2.28$ & $35.31 \pm 2.24$ & $25.05 \pm 7.95 $ \\
			10 &  AE$\oplus$BoF &  $43.34 \pm 2.21$ & $39.15 \pm 1.62$ & $33.14 \pm 6.50 $  \\
			10 &  last$\oplus$BoF &  $\mathbf{44.83 \pm 3.29}$ & $\mathbf{41.81 \pm 4.41}$ & $\mathbf{34.12 \pm 3.72}$ \\
			\hline
		\end{tabular}
	\end{center}
\end{table}

For the hold-out evaluation setup, the performance of the classifier is consistently inferior in comparison to the performance achieved by the SVM. This can be attributed to the fact that these representations are derived in an unsupervised fashion and used as the input to the also unsupervised clustering algorithm, which greatly affects the overall performance of the classifier. In combination with the fact that the hidden layer of the SLFN is not trainable, this constitutes the most major drawback of this classifier. Moreover, the F-score exhibits very high variance in all three prediction targets. This is indicative of the fact that the classifier fails to generalize and make correct predictions about an unseen stock, by using data only from other stocks. Thus, the SLFN fails to capture the general trend of the stock market and apply its knowledge to unknown data.

\subsubsection{MLP Results}

The architecture of the MLP includes an input layer with as many neurons as the dimension of the input representation, two successive hidden layers each with $512$ neurons, and the output layer consisting of three neurons, corresponding to the three possible classes. The results for the anchored day and the hold-out evaluation setups are shown in Table \ref{tab:mlp_day_results} and Table \ref{tab:mlp_stock_results} respectively. The results are notably better than the ones achieved by the previous classifiers. This can be attributed to the fact that MLPs are capable of capturing more complex non-linear relations between the data, are inherently more robust to noisy inputs and can better handle distribution shift phenomena. 

For the anchored day setup, the best performance is achieved when the prediction target is derived as the average movement of the mid price of the next five samples, that is $N_\alpha=5$ and when higher-dimensional representations are used as input, i.e., the combinations of the handcrafted features. The dimensionality of the input data seems to slightly affect the performance of the classifier, although even the low-dimensional representations derived by the AE achieve competitive results. The \emph{concat} and \emph{last$\oplus$mean}  representations, which exhibit the highest dimension amongst the representations, yield the best results.  Moreover, the classifier achieves great performance even when trying to predict the mid price movement of the immediately succeeding $1$ sample.

\begin{table}[h!]
	\footnotesize
	\caption{Anchored walk forward evaluation: MLP}
	\label{tab:mlp_day_results}
	\begin{center}
		\begin{tabular}{rl|ccc}
			$\mathbf{N_\alpha}$ & \normalfont{\textbf{Input}}  &  \textbf{Precision} &  \textbf{Recall } & \textbf{F-score}  \\
			\hline
			1 & last  & $ 57.26 \pm 7.44 $ & $ 37.40 \pm 1.40 $ & $ 38.65 \pm 1.50 $ \\
			1 & mean  & $ 57.61 \pm 6.67 $ & $ 40.94 \pm 1.79 $ & $ 43.63 \pm 2.25 $ \\
			1 & last$\oplus$mean  & $ 62.03 \pm 8.27 $ & $ 43.50 \pm 2.85 $ & $ 46.62 \pm 2.89 $ \\
			1 & concat  & $ \mathbf{62.30 \pm 6.90}$ & $\mathbf{45.40 \pm 3.14}$ & $ \mathbf{48.75 \pm 3.22}$ \\
			\hline
			1 & AE  & $ 45.86 \pm 6.37 $ & $ 33.47 \pm 0.10 $ & $ 31.79 \pm 0.46 $ \\
			
			1 & BoF  & $ 51.53 \pm 7.49 $ & $ 34.43 \pm 0.54 $ & $ 33.77 \pm 0.89 $ \\
			1 & AE$\oplus$BoF  & $ 54.91 \pm 7.41 $ & $ 35.01 \pm 0.74 $ & $ 34.81 \pm 1.17 $ \\
			
			1 & last$\oplus$BoF  & $\mathbf{56.26 \pm 7.50}$ & $\mathbf{38.73 \pm 1.81 }$ & $\mathbf{40.40 \pm 1.47}$ \\
			\hline \hline
			5 & last  & $ 55.24 \pm 5.69 $ & $ 40.82 \pm 2.42 $ & $ 41.98 \pm 2.58 $ \\
			5 & mean  & $ 58.51 \pm 6.33 $ & $ 47.79 \pm 3.14 $ & $ 49.98 \pm 3.23 $ \\
			5 & last$\oplus$mean  & $ 62.67 \pm 6.60 $ & $ 50.26 \pm 3.26 $ & $ 53.06 \pm 3.30 $ \\
			5 & concat  & $ \mathbf{62.71 \pm 5.80}$ & $\mathbf{53.72 \pm 3.63}$ & $ \mathbf{56.06 \pm 2.82}$ \\
			\hline
			5 & AE  & $ 49.41 \pm 7.63 $ & $ 33.69 \pm 0.15 $ & $ 30.11 \pm 1.08 $ \\
			
			5 & BoF & $ 52.79 \pm 5.86 $ & $ 36.47 \pm 0.71 $ & $ 35.78 \pm 1.29 $ \\
			5 & AE$\oplus$BoF & $ 55.03 \pm 5.36 $ & $ 38.05 \pm 1.40 $ & $ 38.32 \pm 2.22 $ \\
			
			5 & last$\oplus$BoF & $ \mathbf{57.18 \pm 6.76}$ & $\mathbf{43.79 \pm 2.64} $ & $\mathbf{45.38 \pm 1.99}$ \\
			\hline \hline
			10 & last & $ 52.17 \pm 4.57 $ & $ 41.91 \pm 2.07 $ & $ 42.87 \pm 2.24 $ \\
			10 & mean & $ 56.32 \pm 5.89 $ & $ 45.06 \pm 2.33 $ & $ 47.08 \pm 2.52 $ \\
			10 & last$\oplus$mean & $ 60.00 \pm 5.96 $ & $ 48.19 \pm 2.17 $ & $ 50.74 \pm 2.43 $ \\
			10 & concat & $ \mathbf{60.61 \pm 6.20} $ & $\mathbf{49.70 \pm 3.17}$ & $ \mathbf{52.03 \pm 3.12}$ \\
			\hline
			10 & AE & $ 49.51 \pm 11.82 $ & $ 33.69 \pm 0.08 $ & $ 29.22 \pm 1.23 $ \\
			
			10 & BoF & $ 51.41 \pm 5.18 $ & $ 37.05 \pm 1.09 $ & $ 36.17 \pm 1.53 $ \\
			10 & AE$\oplus$BoF & $ 53.63 \pm 5.23 $ & $ 38.09 \pm 1.36 $ & $ 37.90 \pm 2.05 $ \\
			
			10 & last$\oplus$BoF & $\mathbf{54.22 \pm 5.89}$ & $\mathbf{43.93 \pm 1.35}$ & $ \mathbf{45.54 \pm 1.51}$ \\
			
			\hline
		\end{tabular}
	\end{center}
\end{table}

\begin{table}[h!]
	\footnotesize
	\caption{Hold-out per stock evaluation: MLP}
	\label{tab:mlp_stock_results}
	\begin{center}
		\begin{tabular}{rl|ccc}
			$\mathbf{N_\alpha}$ & \normalfont{\textbf{Input}}  &  \textbf{Precision} &  \textbf{Recall } & \textbf{F-score}  \\
			\hline
			1 & last & $ 57.43 \pm 2.22 $ & $ 35.61 \pm 0.53 $ & $ 35.30 \pm 1.98 $ \\
			1 & mean & $ 58.28 \pm 5.42 $ & $ 39.32 \pm 4.03 $ & $ 40.05 \pm 4.27 $ \\
			1 & last$\oplus$mean  & $ 59.15 \pm 5.84 $ & $ 42.38 \pm 5.52 $ & $ 43.08 \pm 5.07 $ \\
			1 & concat  & $ \mathbf{60.30 \pm 4.96}$ & $\mathbf{43.11 \pm 2.01}$ & $ \mathbf{46.05 \pm 2.55}$ \\
			\hline
			1 & AE  & $ 40.93 \pm 5.73 $ & $ 33.50 \pm 0.16 $ & $ 31.26 \pm 1.54 $ \\
			
			1 & BoF & $ 45.52 \pm 3.77 $ & $ 34.07 \pm 0.58 $ & $ 32.31 \pm 2.62 $ \\
			1 & AE$\oplus$BoF & $ 48.93 \pm 1.96 $ & $ 34.20 \pm 0.52 $ & $ 32.62 \pm 2.02 $ \\
			
			1 & last$\oplus$BoF & $\mathbf{53.19 \pm 7.01}$ & $\mathbf{37.79 \pm 2.48}$ & $\mathbf{38.37 \pm 3.90}$ \\
			\hline \hline
			5 & last & $ 54.88 \pm 5.36 $ & $ 37.86 \pm 2.28 $ & $ 36.96 \pm 3.11 $ \\
			5 & mean & $ 60.08 \pm 1.36 $ & $ 44.96 \pm 2.21 $ & $ 47.03 \pm 3.39 $ \\
			5 & last$\oplus$mean & $\mathbf{62.53 \pm 2.22}$ & $ 49.62 \pm 2.93 $ & $ \mathbf{52.12 \pm 3.20}$ \\
			5 & concat & $ 61.29 \pm 6.83 $ & $\mathbf{52.48 \pm 5.88}$ & $ 51.90 \pm 5.83 $ \\
			\hline
			5 & AE & $ 40.32 \pm 7.26 $ & $ 33.48 \pm 0.11 $ & $ 28.82 \pm 1.86 $ \\
			
			5 & BoF & $ 47.91 \pm 3.43 $ & $ 35.63 \pm 1.12 $ & $ 33.31 \pm 3.42 $ \\
			5 & AE$\oplus$BoF & $ 52.85 \pm 1.41 $ & $ 36.25 \pm 1.89 $ & $ 34.41 \pm 4.00 $ \\
			
			5 & last$\oplus$BoF & $\mathbf{58.07 \pm 7.30}$ & $\mathbf{39.31 \pm 4.36}$ & $\mathbf{38.23 \pm 6.54}$ \\
			\hline \hline
			10 & last & $ 52.80 \pm 4.79 $ & $ 39.34 \pm 3.38 $ & $ 37.61 \pm 5.99 $ \\
			10 & mean & $ 58.30 \pm 2.23 $ & $ 45.31 \pm 2.73 $ & $ 47.25 \pm 2.84 $ \\
			10 & last$\oplus$mean & $\mathbf{62.44 \pm 2.59}$ & $ 44.11 \pm 2.82 $ & $ 45.44 \pm 2.71 $ \\
			10 & concat & $ 61.19 \pm 2.71 $ & $\mathbf{48.68 \pm 4.23}$ & $\mathbf{50.36 \pm 5.47}$ \\
			\hline
			10 & AE & $ 41.96 \pm 6.15 $ & $ 33.45 \pm 0.06 $ & $ 27.66 \pm 2.31 $ \\
			
			10 & BoF & $ 48.57 \pm 1.51 $ & $ 35.86 \pm 1.89 $ & $ 32.87 \pm 4.59 $ \\
			10 & AE$\oplus$BoF & $ 51.22 \pm 2.37 $ & $ 36.60 \pm 2.27 $ & $ 34.10 \pm 4.81 $ \\
			
			10 & last$\oplus$BoF & $\mathbf{53.61 \pm 5.59}$ & $\mathbf{41.05 \pm 3.42} $ & $\mathbf{40.60 \pm 4.85}$ \\
			
			\hline
		\end{tabular}
	\end{center}
\end{table}	

For the hold-out setup, the classifier seems capable of making correct predictions for all three sets of labels evaluated, i.e., for $N_\alpha=1, 5\text{ and }10$. In fact, the MLP achieves the best performance of all three classifiers in this setup, meaning that it is capable of making better generalizations about unknown stock data, by learning from other stocks. Also, as dimensionality increases so does the performance of the classifier.

%R1C3
Furthermore, to provide better insight on the performance of the best model (MLP) we report the evaluated metrics separately for each class using the best available representation (\emph{concat}) for the $N_a=10$ scenario, which a trader might be most interested in. The results are summarized in Table~\ref{table:per-class-scores}. The results, in particular the precision scores, indicate that the classifier is capable of making a correct decision for the up and down classes at a rate higher than random guessing.  However note that the recall and, by extension, the F-scores are highly affected by the very large number of samples in our dataset and don't necessarily reflect the requirements a trader might expect from a classifier, as traders act on positive signals. The lower performance of the classifier for short-term forecasts can be also attributed to the noisy nature of the mid-price at very short prediction horizons --- the model performs significantly better for predicting the longer term behavior of the stocks.

\begin{table}[!ht]
	\caption{Per class scores achieved by the MLP classifier using the \emph{concat} representation.}
	\label{table:per-class-scores}
	\begin{center}
		\begin{tabular}{ cccc }
			\textbf{Class} & \textbf{Precision} & \textbf{Recall} & \textbf{F-score} \\  \hline
			$\uparrow$ & $ 49.60 \pm 9.68 $ & $ 32.40 \pm 7.44 $ & $ 37.96 \pm 4.52 $ \\
			--- & $ 79.35 \pm 5.25 $ & $ 89.47 \pm 6.91 $ & $ 93.98 \pm 5.04 $ \\
			$\downarrow$ & $ 52.90 \pm 11.33 $ & $ 27.23 \pm 8.68 $ & $ 34.14 \pm 5.78 $ \\ \hline 
		\end{tabular}
	\end{center}
\end{table}

%R3C3
Finally, to examine the effect of the used MLP architecture on the quality of the learned model, we evaluated the MLP using a number of different architectures by varying the number of hidden layers and neurons per layer. The results are reported in Table~\ref{table:mlp-sizes}. Only the evaluation results for the anchored walk forward setup using the \emph{concat} representation are reported due to lack of space. However, similar results were obtained for the rest of the representations and evaluation setups. Even though the used ``512-512-3'' architecture leads to the best F-score, using a different architecture does not severely impact the quality of the learned model.

\begin{table}[h!]
	\scriptsize
	\caption{Comparing different MLP architectures using the \emph{concat} representation (anchored walk forward setup, $N_\alpha=10$).}
	\label{table:mlp-sizes}
	\begin{center}
		\begin{tabular}{lcccc}
			\textbf{Hidden Layers} &  \textbf{Precision} &  \textbf{Recall } & \textbf{F-score}  \\
			\hline
			512 & $ 61.24 \pm 6.29 $ & $ 43.69 \pm 2.85 $ & $ 47.03 \pm 3.25 $ \\
			1024 & $ 61.09 \pm 6.18 $ & $43.51 \pm 2.07 $ & $ 46.97 \pm 1.96 $\\
			512-512 & $\mathbf{62.30 \pm 7.18}$ & $\mathbf{45.40 \pm 3.27}$ & $\mathbf{48.75 \pm 3.36}$ \\		
			1024-1024-1024 & $ 61.99 \pm 7.98 $ & $ 44.65 \pm 4.92 $ & $ 46.98 \pm 3.57 $ \\			
			\hline
		\end{tabular}
	\end{center}
\end{table}

The activations of the hidden layers of the MLP can be thought of as features learned from the input representation, which are biased towards the task of classification. Thus, the MLP performs a kind of supervised feature extraction, where the extracted features are learned via optimizing the classification error. On the contrary, the AE and BoF models perform unsupervised feature extraction, which has no guarantee of being suitable for the task at hand. The classification-biased feature learning process that occurs in parallel with the classifier's training, has the potential to produce very robust features and lead to improved predictions. This is reflected by the slightly deteriorated performance of the classifier, when the input representation is derived by unsupervised feature extraction techniques.   In combination with the supervised feature learning performed by training the MLP, this further reaffirms the assumption that the activations of the MLP's hidden layers learn representations of the data which highly befit the task of classification.

%R1C2
%
\subsection{Computational Complexity Analysis}

Financial exchanges generate a vast amount of data that must be processed in real-time in order to quickly respond to the volatile conditions of the markets. Therefore, the speed of the deployed models is equally important with the forecasting accuracy in real-word applications where large amounts of data must be processed under strict time constrains. In this Subsection we provide both an asymptotic forecasting time analysis and an empirical study of the run time of the used models. Our analysis is focused on the prediction complexity, since the real-time constrains can be relaxed during training by using a slightly outdated model. Furthermore, all the used models can be incrementally trained and, as a result, adapt to the available computational resources during the training.

Let $N_r$ be the size of the used representation, $N_C$ be the number of possible mid-price movements ($N_C=3$) and $N_h$ be the size of the hidden representation (used only for the SLFN and the MLP models). Table~\ref{table:computational-complexity} summarizes the time complexity of the models. The SVM has significantly lower time requirements than the other two models. The complexity of the SLFN and MLP models can also be adjusted by altering the size of the hidden layers ($N_h=1000$ for the SLFN and $N_h=512$ for the MLP).  Note that the used SVM operates in the primal space and therefore the time complexity does not depend on the number of selected support vectors.

The results of the empirical run time analysis, reported in Table 10, also confirm the previous findings. All the classifiers	were implemented using GPU-accelerated libraries [45] and a mid-range GPU with 6GB of RAM was used for the conducted experiments. To ensure a fair comparison of the compared classifiers we report the time needed after the feature extraction step. Note that even the MLP classifier, which is the most computationally-intensive model, is able to process 	almost 7,000 transactions per second using a mid-range GPU. On the other hand, the SVM achieves the best trade-off between forecasting accuracy and run time, being capable of processing more than 14,000 transactions per second.

\begin{table}[!ht]
	\caption{Computational complexity analysis of the used classifiers}
	\label{table:computational-complexity}
	\begin{center}
		\begin{tabular}{ lcc }		\textbf{Model} & \textbf{Time Complexity} \\  \hline
			{SVM} & $N_r N_C$ \\  
			{SLFN} & $N_r N_h + N_h N_C$  \\
			{MLP} & $N_r N_h + N_h^2+ N_h N_C$  \\
		\end{tabular}
	\end{center}
\end{table}

\begin{table}[!ht]
	\caption{Run time analysis of the used classifiers. The mean time observed over 1000 runs measured in milliseconds is reported.}
	\label{table:runtime-analysis}
	\begin{center}
		\begin{tabular}{ lccc }	
			\textbf{Representation}  & \textbf{SVM} & \textbf{SLFN} & \textbf{MLP} \\  \hline
			{AE} & $0.062$ ms & $0.090$ ms  & $0.124$ ms\\
			{BoF} & $0.063$ ms & $0.093$ ms  & $0.127$ ms\\
			{last} &  $0.065$ ms & $0.093$ ms  & $0.133$ ms\\
			{mean} & $0.065$ ms & $0.093$ ms  & $0.133$ ms\\
			{AE$\oplus$BoF}  & $0.065$ ms & $0.093$ ms  & $0.130$ ms\\
			{last$\oplus$BoF}  & $0.065$ ms & $0.093$ ms  & $0.138$ ms\\
			{last$\oplus$mean}  & $0.065$ ms & $0.098$ ms  & $0.136$ ms\\
			{concat}  & $0.068$ ms & $0.112$ ms  & $0.144$ ms\\	 
		\end{tabular}
	\end{center}
\end{table}

%R3C4
\subsection{Statistical Analysis}

\begin{figure}[!ht]
	\begin{center}
		\includegraphics[trim={5cm 14cm 5cm 9cm}, scale=0.7,clip ]{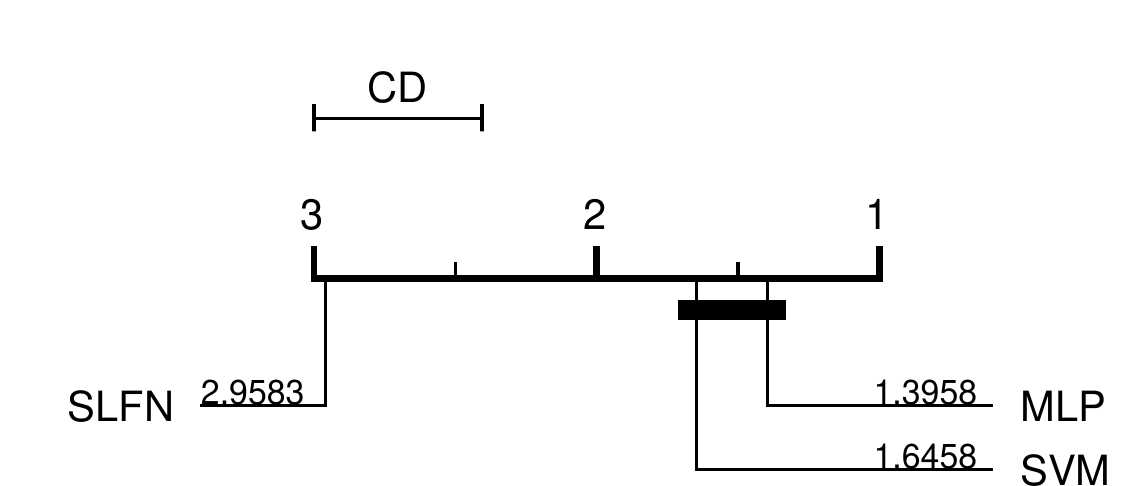}
	\end{center}
	\caption{Nemenyi post-hoc test: Comparing different classifiers}
	\label{fig:nemeyi-classifiers}
\end{figure}

To validate the significance of the obtained results we performed a series of statistical tests. First, the three different classifiers used for forecasting the mid price movements were compared using the Friedman's test~\cite{hollander2013nonparametric}. The null hypothesis was defined as: ``\textit{There is no statistical significant difference ($a=0.1$) between the SVM, SLFN and MLP classifiers.}'' To compare the classifiers the F-score for the different representations and predictions horizons was used. The null hypothesis was rejected ($p=3.9\times 10^{-26}$), meaning that at least one classifier is significantly better than the others. The Nemenyi post-hoc test ~\cite{hollander2013nonparametric} was then used to evaluate the differences between the classifiers, as shown in Figure~\ref{fig:nemeyi-classifiers}. The SVM and MLP classifiers are significantly better than the SLFN classifier. Note that even though the MLP classifier performs better than the SVM, the differences between the MLP and the SVM are not statistically significant ($a=0.1$).

\begin{figure}[!ht]
	\begin{center}
		\includegraphics[trim={4.8cm 11.5cm 4.2cm 9cm}, scale=0.65,clip ]{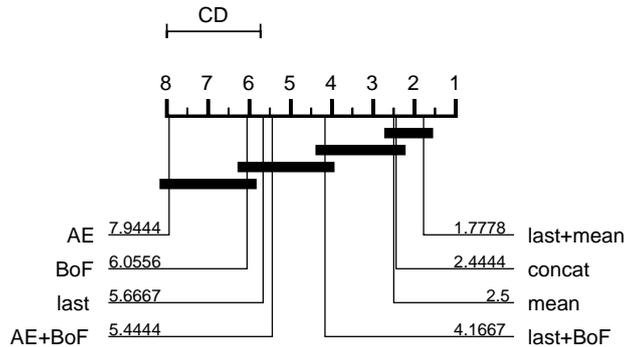}
	\end{center}
	\caption{Nemenyi post-hoc test: Comparing different representations}
	\label{fig:nemeyi-representations}	
\end{figure}

We also compared the eight different representations used for the conducted experiments. The null hypothesis was defined as: \textit{``There is no statistical significant difference ($a=0.1$) between the eight used representations.''} Again, the null hypothesis was strongly rejected using the Friedman's test ($p=5.5\times 10^{-35}$). To further compare the representations the Nemenyi post-hoc test was used as before. The results are shown in Figure~\ref{fig:nemeyi-representations}. The \emph{concat}, \emph{last+mean} and \emph{mean} representations are significantly better than most of the other used representations. Also, note that the dimensionality of a representation seems to be correlated with its predictive power. However, even though the \emph{last} and \emph{mean} representations have the same dimensionality, the \emph{mean} representation leads to significantly better results. This is mainly due to the de-noising effect of the averaging process used for extracting the \emph{mean} representation, effectively suppressing possible outliers.

\subsection{Evaluation Summary}

As corroborated by the statistical tests performed, the handcrafted representations and especially those that incorporate temporal information yield the most significant results. The \emph{BoF} and \emph{last$\oplus$BoF} representations follow, as they also take into consideration past values during the feature extraction process, but as the input dimension decreases so does the discriminative ability of the classifiers. The MLP classifier performs better when the input representation is derived by the handcrafted features, as its hidden layers extract a classification-based representation of its input. The low dimensionality of the \emph{AE}, and \emph{BoF} representations in combination with their unsupervised nature lead the MLP to make fewer correct predictions.
The low dimensionality of these learned representations allows for faster computations, albeit at the cost of achieving slightly deteriorated performances. The SVM classifier's performance increases with the dimensionality of the input representation, but falls behind the MLP's performance, as it is less resistant to the inherent noise of the data. Finally, the SLFN's performance is plagued by the high number of false negative predictions. The incorporation of past information does somewhat alleviate this drawback, but its overall performance falls back in comparison to the other classifiers. 

\section{Conclusions and Discussion}
\label{sec:conclusion}

In this work, an extensive study into the information provided by the high frequency limit order book with respect to the forecasting of future mid price movements was presented. Several representations derived by handcrafted features as well as features learned by Machine Learning algorithms, ranging from 24 to 720-dimensional feature vectors, were considered and used as input to various classifiers for the forecasting task. Three scenarios were assessed regarding the span of time for which predictions are made. Finally, two evaluation analysis methods were examined for each classifier, scenario and input representation. Through the anchored walk forward setup, the ability of the evaluated classifiers to learn from past stock data and apply this knowledge to future, unknown data is determined. The hold-out setup serves to examine whether the classifiers are able to capture the general trends and movements of the stock market by learning from some stocks and applying this knowledge to unseen stocks.

The results achieved are remarkable in all cases, indicating that Machine Learning techniques are capable of correctly predicting mid price movements. Different classifiers seem to perform better in different aspects, such as the precision or recall of the predicted movements. We have provided the first, to the best of our knowledge, in depth review and evaluation that addresses the challenging characteristics of LOB data, such as the high velocity, variance, volume and strict real-time constraints, and uses ML techniques to predict the mid price movement, providing insight into the information contained in a LOB. The prediction results are improved when combining the extracted feature representations with the handcrafted ones, indicating that the feature extraction models are able to uncover latent, auxiliary knowledge. Finally, the learned representations also yield significant results when used alone and systematically improve the time-wise performance of all classifiers by reducing the dimensionality of the input data.

%\section*{References}
\bibliographystyle{elsarticle-num}

\begin{thebibliography}{10}
	\expandafter\ifx\csname url\endcsname\relax
	\def\url#1{\texttt{#1}}\fi
	\expandafter\ifx\csname urlprefix\endcsname\relax\def\urlprefix{URL }\fi
	\expandafter\ifx\csname href\endcsname\relax
	\def\href#1#2{#2} \def\path#1{#1}\fi
	

\bibitem{kaastra1995forecasting}
I.~Kaastra and M.~S. Boyd, ``Forecasting futures trading volume using neural
  networks,'' \emph{Journal of Futures Markets}, vol.~15, no.~8, pp. 953--970,
  1995.

\bibitem{kaastra1996designing}
------, ``Designing a neural network for forecasting financial and economic
  time series,'' \emph{Neurocomputing}, vol.~10, no.~3, pp. 215--236, 1996.

\bibitem{tay2001application}
F.~E. Tay and L.~Cao, ``Application of support vector machines in financial
  time series forecasting,'' \emph{Omega}, vol.~29, no.~4, pp. 309--317, 2001.

\bibitem{cao2003support}
L.-J. Cao and F.~E.~H. Tay, ``Support vector machine with adaptive parameters
  in financial time series forecasting,'' \emph{IEEE Transactions on Neural
  Networks}, vol.~14, no.~6, pp. 1506--1518, 2003.

\bibitem{lu2009financial}
C.-J. Lu, T.-S. Lee, and C.-C. Chiu, ``Financial time series forecasting using
  independent component analysis and support vector regression,''
  \emph{Decision Support Systems}, vol.~47, no.~2, pp. 115--125, 2009.

\bibitem{srivastava2014dropout}
N.~Srivastava, G.~E. Hinton, A.~Krizhevsky, I.~Sutskever, and R.~Salakhutdinov,
  ``Dropout: a simple way to prevent neural networks from overfitting.''
  \emph{Journal of Machine Learning Research}, vol.~15, no.~1, pp. 1929--1958,
  2014.

\bibitem{glorot2010understanding}
X.~Glorot and Y.~Bengio, ``Understanding the difficulty of training deep
  feedforward neural networks,'' in \emph{Proceedings of the thirteenth
  international conference on artificial intelligence and statistics}, 2010,
  pp. 249--256.

\bibitem{langkvist2014review}
M.~L{\"a}ngkvist, L.~Karlsson, and A.~Loutfi, ``A review of unsupervised
  feature learning and deep learning for time-series modeling,'' \emph{Pattern
  Recognition Letters}, vol.~42, pp. 11--24, 2014.

\bibitem{rahman2016layered}
M.~M. Rahman, M.~M. Islam, K.~Murase, and X.~Yao, ``Layered ensemble
  architecture for time series forecasting,'' \emph{IEEE transactions on
  cybernetics}, vol.~46, no.~1, pp. 270--283, 2016.

\bibitem{vincent2008extracting}
P.~Vincent, H.~Larochelle, Y.~Bengio, and P.-A. Manzagol, ``Extracting and
  composing robust features with denoising autoencoders,'' in \emph{Proceedings
  of the International Conference on Machine Learning}, 2008, pp. 1096--1103.

\bibitem{baydogan2013bag}
M.~G. Baydogan, G.~Runger, and E.~Tuv, ``A bag-of-features framework to
  classify time series,'' \emph{IEEE Transactions on Pattern Analysis and
  Machine Intelligence}, vol.~35, no.~11, pp. 2796--2802, 2013.

\bibitem{iosifidis2013multidimensional}
A.~Iosifidis, A.~Tefas, and I.~Pitas, ``Multidimensional sequence
  classification based on fuzzy distances and discriminant analysis,''
  \emph{IEEE Transactions on Knowledge and Data Engineering}, vol.~25, no.~11,
  pp. 2564--2575, 2013.

\bibitem{park2015using}
B.~Park and J.~K. Bae, ``Using machine learning algorithms for housing price
  prediction: The case of fairfax county, virginia housing data,'' \emph{Expert
  Systems with Applications}, vol.~42, no.~6, pp. 2928--2934, 2015.

\bibitem{yu2008forecasting}
L.~Yu, S.~Wang, and K.~K. Lai, ``Forecasting crude oil price with an emd-based
  neural network ensemble learning paradigm,'' \emph{Energy Economics},
  vol.~30, no.~5, pp. 2623--2635, 2008.

\bibitem{kercheval2015modelling}
A.~N. Kercheval and Y.~Zhang, ``Modelling high-frequency limit order book
  dynamics with support vector machines,'' \emph{Quantitative Finance},
  vol.~15, no.~8, pp. 1315--1329, 2015.

\bibitem{cont2011statistical}
R.~Cont, ``Statistical modeling of high-frequency financial data,'' \emph{IEEE
  Signal Processing Magazine}, vol.~28, no.~5, pp. 16--25, 2011.

\bibitem{engle2004analysis}
R.~F. Engle and J.~R. Russell, ``Analysis of high frequency financial data,''
  \emph{Handbook of financial econometrics}, 2004.

\bibitem{bouchaud2002statistical}
J.-P. Bouchaud, M.~M{\'e}zard, and M.~Potters, ``Statistical properties of
  stock order books: empirical results and models,'' \emph{Quantitative
  Finance}, vol.~2, no.~4, pp. 251--256, 2002.

\bibitem{huang2005forecasting}
W.~Huang, Y.~Nakamori, and S.-Y. Wang, ``Forecasting stock market movement
  direction with support vector machine,'' \emph{Computers \& Operations
  Research}, vol.~32, no.~10, pp. 2513--2522, 2005.

\bibitem{kim2003financial}
K.-j. Kim, ``Financial time series forecasting using support vector machines,''
  \emph{Neurocomputing}, vol.~55, no.~1, pp. 307--319, 2003.

\bibitem{zhang2001multiresolution}
B.-L. Zhang, R.~Coggins, M.~A. Jabri, D.~Dersch, and B.~Flower,
  ``Multiresolution forecasting for futures trading using wavelet
  decompositions,'' \emph{IEEE Transactions on Neural Networks}, vol.~12,
  no.~4, pp. 765--775, 2001.

\bibitem{tran2019data}
D.~T. Tran, J.~Kanniainen, M.~Gabbouj, and A.~Iosifidis, ``Data-driven neural
  architecture learning for financial time-series forecasting,''
  \emph{arXiv:1903.06751}, 2019.

\bibitem{tran2017tensor}
D.~T. Tran, M.~Magris, J.~Kanniainen, M.~Gabbouj, and A.~Iosifidis, ``Tensor
  representation in high-frequency financial data for price change
  prediction,'' in \emph{2017 IEEE Symposium Series on Computational
  Intelligence}.\hskip 1em plus 0.5em minus 0.4em\relax IEEE, 2017, pp. 1--7.

\bibitem{tran2018temporal}
D.~T. Tran, A.~Iosifidis, J.~Kanniainen, and M.~Gabbouj, ``Temporal
  attention-augmented bilinear network for financial time-series data
  analysis,'' \emph{IEEE Transactions on Neural Networks and Learning Systems,
  DOI: 10.1109/TNNLS.2018.2869225}, 2018.

\bibitem{walczak2001empirical}
S.~Walczak, ``An empirical analysis of data requirements for financial
  forecasting with neural networks,'' \emph{Journal of Management Information
  Systems}, vol.~17, no.~4, pp. 203--222, 2001.

\bibitem{heaton2016deep}
J.~Heaton, N.~Polson, and J.~Witte, ``Deep portfolio theory,'' \emph{arXiv
  preprint arXiv:1605.07230}, 2016.

\bibitem{tsantekidis2017using}
A.~Tsantekidis, N.~Passalis, A.~Tefas, J.~Kanniainen, M.~Gabbouj, and
  A.~Iosifidis, ``Using deep learning to detect price change indications in
  financial markets,'' in \emph{Proceedings of the European Signal Processing
  Conference}, 2017, pp. 2511--2515.

\bibitem{tsantekidis2017forecasting}
------, ``Forecasting stock prices from the limit order book using
  convolutional neural networks,'' in \emph{Proceedings of the IEEE Conference
  on Business Informatics}, vol.~1, 2017, pp. 7--12.

\bibitem{passalis2017time}
N.~Passalis, A.~Tsantekidis, A.~Tefas, J.~Kanniainen, M.~Gabbouj, and
  A.~Iosifidis, ``Time-series classification using neural bag-of-features,'' in
  \emph{Proceedings of the European Signal Processing Conference}, 2017, pp.
  301--305.

\bibitem{siikanen2017limit}
M.~Siikanen, J.~Kanniainen, and J.~Valli, ``Limit order books and liquidity
  around scheduled and non-scheduled announcements: Empirical evidence from
  nasdaq nordic,'' \emph{Finance Research Letters}, vol.~21, pp. 264--271,
  2017.

\bibitem{siikanen2017drives}
M.~Siikanen, J.~Kanniainen, and A.~Luoma, ``What drives the sensitivity of
  limit order books to company announcement arrivals?'' \emph{Economics
  Letters}, vol. 159, pp. 65--68, 2017.

\bibitem{ntakaris2017benchmark}
A.~Ntakaris, M.~Magris, J.~Kanniainen, M.~Gabbouj, and A.~Iosifidis,
  ``Benchmark dataset for mid-price forecasting of limit order book data with
  machine learning methods,'' \emph{Journal of Forecasting, DOI:
  10.1002/for.2543}, 2018.

\bibitem{galar2012review}
M.~Galar, A.~Fernandez, E.~Barrenechea, H.~Bustince, and F.~Herrera, ``A review
  on ensembles for the class imbalance problem: bagging-, boosting-, and
  hybrid-based approaches,'' \emph{IEEE Transactions on Systems, Man, and
  Cybernetics, Part C (Applications and Reviews)}, vol.~42, no.~4, pp.
  463--484, 2012.

\bibitem{longadge2013class}
R.~Longadge and S.~Dongre, ``Class imbalance problem in data mining review,''
  \emph{arXiv preprint arXiv:1305.1707}, 2013.

\bibitem{hinton2006reducing}
G.~E. Hinton and R.~R. Salakhutdinov, ``Reducing the dimensionality of data
  with neural networks,'' \emph{Science}, vol. 313, no. 5786, pp. 504--507,
  2006.

\bibitem{ngiam2011multimodal}
J.~Ngiam, A.~Khosla, M.~Kim, J.~Nam, H.~Lee, and A.~Y. Ng, ``Multimodal deep
  learning,'' in \emph{Proceedings of the International Conference on Machine
  Learning}, 2011, pp. 689--696.

\bibitem{rumelhart1988learning}
D.~E. Rumelhart, G.~E. Hinton, and R.~J. Williams, ``Learning representations
  by back-propagating errors,'' \emph{Cognitive modeling}, vol.~5, no.~3, p.~1,
  1988.

\bibitem{zhang2004solving}
T.~Zhang, ``Solving large scale linear prediction problems using stochastic
  gradient descent algorithms,'' in \emph{Proceedings of the International
  Conference on Machine Learning}, 2004, p. 116.

\bibitem{sivic2003video}
J.~Sivic, A.~Zisserman \emph{et~al.}, ``Video google: A text retrieval approach
  to object matching in videos.'' in \emph{Proceedings of the International
  Conference on Computer Vision}, vol.~2, 2003, pp. 1470--1477.

\bibitem{passalis2016information}
N.~Passalis and A.~Tefas, ``Information clustering using manifold-based
  optimization of the bag-of-features representation,'' \emph{IEEE transactions
  on cybernetics}, vol.~48, no.~1, pp. 52--63, 2018.

\bibitem{salton1983mcgill}
G.~Salton and J.~Michael, ``Mcgill,'' \emph{Introduction to modern information
  retrieval}, pp. 24--51, 1983.

\bibitem{kanungo2002efficient}
T.~Kanungo, D.~M. Mount, N.~S. Netanyahu, C.~D. Piatko, R.~Silverman, and A.~Y.
  Wu, ``An efficient k-means clustering algorithm: Analysis and
  implementation,'' \emph{IEEE Transactions on Pattern Analysis and Machine
  Intelligence}, vol.~24, no.~7, pp. 881--892, 2002.

\bibitem{nbof}
N.~Passalis and A.~Tefas, ``Neural bag-of-features learning,'' \emph{Pattern
  Recognition}, vol.~64, pp. 277--294, Apr. 2017.

\bibitem{bousquet2008tradeoffs}
O.~Bousquet and L.~Bottou, ``The tradeoffs of large scale learning,'' in
  \emph{Proceedings of the Advances in Neural Information Processing Systems},
  2008, pp. 161--168.

\bibitem{hoi2009semisupervised}
S.~C. Hoi, R.~Jin, J.~Zhu, and M.~R. Lyu, ``Semisupervised svm batch mode
  active learning with applications to image retrieval,'' \emph{ACM
  Transactions on Information Systems}, vol.~27, no.~3, p.~16, 2009.

\bibitem{hofmann2008kernel}
T.~Hofmann, B.~Sch{\"o}lkopf, and A.~J. Smola, ``Kernel methods in machine
  learning1,'' \emph{The Annals of Statistics}, vol.~36, no.~3, pp. 1171--1220,
  2008.

\bibitem{robert2014machine}
C.~Robert, ``Machine learning, a probabilistic perspective,'' 2014.

\bibitem{zhang2009prototype}
K.~Zhang, J.~T. Kwok, and B.~Parvin, ``Prototype vector machine for large scale
  semi-supervised learning,'' in \emph{Proceedings of the International
  Conference on Machine Learning}, 2009, pp. 1233--1240.

\bibitem{Iosifidis2017210}
A.~Iosifidis, A.~Tefas, and I.~Pitas, ``Approximate kernel extreme learning
  machine for large scale data classification,'' \emph{Neurocomputing}, vol.
  219, pp. 210 -- 220, 2017.

\bibitem{iosifidis2015sparse}
------, ``Sparse extreme learning machine classifier exploiting intrinsic
  graphs,'' \emph{Pattern Recognition Letters}, vol.~65, pp. 192--196, 2015.

\bibitem{haykin2009neural}
S.~S. Haykin, \emph{Neural networks and learning machines}.\hskip 1em plus
  0.5em minus 0.4em\relax Pearson Upper Saddle River, USA, 2009, vol.~3.

\bibitem{kingma2014adam}
D.~Kingma and J.~Ba, ``Adam: A method for stochastic optimization,''
  \emph{arXiv preprint arXiv:1412.6980}, 2014.

\bibitem{tomasini2011trading}
E.~Tomasini and U.~Jaekle, \emph{Trading Systems}.\hskip 1em plus 0.5em minus
  0.4em\relax Harriman House Limited, 2011.

\bibitem{hollander2013nonparametric}
M.~Hollander, D.~A. Wolfe, and E.~Chicken, \emph{Nonparametric statistical
  methods}.\hskip 1em plus 0.5em minus 0.4em\relax John Wiley \& Sons, 2013.

	
\end{thebibliography}

\end{document}